\documentclass[sigconf, screen]{acmart}
\PassOptionsToPackage{dvipsnames}{xcolor}

\PassOptionsToPackage{colorlinks=true,linkcolor=RubineRed,breaklinks=True,citecolor=RubineRed}{xcolor}

\usepackage{framed}

\usepackage{enumitem}
\setlist[itemize]{noitemsep, topsep=0pt}

\usepackage[most]{tcolorbox}
\usepackage{adjustbox}
\usepackage[linesnumbered,algo2e,ruled]{algorithm2e}
\usepackage{multirow}
\usepackage{booktabs,subcaption,dcolumn}
\usepackage{tikz}
\usepackage{enumitem}
\usepackage{tabularx}
\usepackage{MnSymbol}

\usepackage{pifont}
\usepackage{svg}
\usepackage{soul}
\usepackage{amsmath}
\usepackage{dirtytalk}
\usepackage{subcaption}
\usepackage{graphicx}
\usepackage{setspace}
\usepackage[font=small]{caption}
\usepackage{booktabs} 
\usepackage[titletoc,toc,title]{appendix}
\usepackage{svg}
\usepackage{amsfonts}
\usepackage{xurl} 

\usepackage{diagbox}
\usepackage{indentfirst}

\usepackage{CJKutf8}
\usepackage{amsmath}
\usepackage{amsfonts} 

\usepackage{colortbl}

\usepackage{siunitx}
\usepackage{setspace}

\AtBeginDocument{%
  \providecommand\BibTeX{{%
    \normalfont B\kern-0.5em{\scshape i\kern-0.25em b}\kern-0.8em\TeX}}}

\acmYear{2026}\copyrightyear{2026}
\setcopyright{cc}
\setcctype[4.0]{by-nc-nd}
\acmConference[ASIA CCS '26]{ACM Asia Conference on Computer and Communications Security}{June 1--5, 2026}{Bangalore, India}
\acmBooktitle{ACM Asia Conference on Computer and Communications Security (ASIA CCS '26), June 1--5, 2026, Bangalore, India}
\acmDOI{10.1145/3779208.3807482}
\acmISBN{979-8-4007-2356-8/26/06}

\newcolumntype{?}{!{\vrule width 1.5pt}}

\newcommand{\concap}[1]{{{\fontfamily{qcr}\selectfont C\MakeLowercase{on}C\MakeLowercase{ap}}}}
\newcommand{\command}[2]{{#2{\fontfamily{qhv}\selectfont #1}}}

\newcommand{\textbox}[1]{
    \noindent\fbox{%
        \parbox{0.97\columnwidth}{%
            {#1}
        }%
    }
}

\newcommand\smamath[1]{{\small $#1$}}
\newcommand\smacal[1]{{\small $\mathcal{#1}$}}

\newcommand\scmath[1]{{\scriptsize $#1$}}

\input{auxiliary/acm_variant}
\begin{document}
\title{``What is the Problem Space?'' Defining Host-space Adversarial Perturbations against Network Intrusion Detection Systems}
\author{Miel Verkerken}

\authornote{Please contact Miel Verkerken for inquiries also at {mielverkerken@hotmail.com}}
\orcid{0000-0002-1781-900X}
\affiliation{%
  \institution{Ghent University - imec}
  \city{Ghent}
  \country{Belgium}}
\email{miel.verkerken@ugent.be}

\author{Laurens D'hooge}
\orcid{0000-0001-5086-6361}
\affiliation{%
  \institution{Ghent University - imec}
  \city{Ghent}
  \country{Belgium}}
\email{laurens.dhooge@ugent.be}

\author{Bruno Volckaert}
\orcid{0000-0003-0575-5894}
\affiliation{%
  \institution{Ghent University - imec}
  \city{Ghent}
  \country{Belgium}}
\email{bruno.volckaert@ugent.be}

\author{Filip De Turck}
\orcid{0000-0003-4824-1199}
\affiliation{%
  \institution{Ghent University - imec}
  \city{Ghent}
  \country{Belgium}}
\email{filip.deturck@ugent.be}

\author{Giovanni Apruzzese}
\orcid{0000-0002-6890-9611}
\affiliation{%
  \institution{University of Liechtenstein}
  \city{Vaduz}
  \country{Liechtenstein}}
\affiliation{%
  \institution{Reykjavik University}
  \city{Reykjavik}
  \country{Iceland}}
\email{giovannia@ru.is}


\begin{abstract}

\noindent
Network Intrusion Detection Systems (NIDS) are now increasingly leveraging Machine Learning (ML) techniques to detect malicious network activities. 
Numerous papers have scrutinized the security of ML-based NIDS (ML-NIDS) by testing them against various attacks involving adversarial perturbations. The findings were oftentimes worrying: by making imperceptible changes to a given input, powerful ML models would be bypassed. In this context, we took a step back and wondered: where (i.e., in what ``space'') have these perturbations been applied?

We argue that real-world adversaries can apply adversarial perturbations only by operating on the hosts they can control---a concept which we define as \textit{host-space perturbations}. To some, such an observation may seem trivial. And yet, through a systematic literature review (n=316), we found that prior work applied perturbations by manipulating pre-collected datapoints (e.g., a packet \textit{captured by the router}, or a network flow \textit{analysed by the ML-NIDS}). Such operations, while not impossible, may be outside the reach of an attacker who can only control some (unprivileged) hosts in a network. Hence, to demonstrate how to craft host-space perturbations and study some of their effects, we experimented on well-known benchmarks and a real-world network. We show that ML-NIDS that can detect the SSH-bruteforcing attempts launched via a given command string cannot detect any attempt launched by changing \textit{a single character} of such a string.
We then examined how such a minuscule change in the ``problem space'' (i.e., the attacker's host) can lead to devastating effects on the ``feature space''. We derive lessons learned on how to practically assess host-space perturbations. Our stance is that the security of ML-NIDS should be re-assessed. 
\end{abstract}



\settopmatter{printfolios=true}

\maketitle

\section{Introduction}
\label{sec:introduction}

\noindent
Assessing the security of Machine Learning (ML) systems is challenging. On the one hand, researchers must foresee the various ways in which a hypothetical attacker can interact with the envisioned ML system~\cite{biggio2018wild}. On the other hand, such interactions must be faithfully reproduced in experiments that resemble a real-world setup~\cite{pierazzi2020intriguing}. Achieving both in the context of modern networks, known for their immense variability~\cite{sommer2010outside}, is objectively hard.

Yet, in the specific domain of Network Intrusion Detection Systems (NIDS), there is a never-ending need of such assessments. Real-world evidence shows that ML models are no longer a ``research toy''. Rather, they are increasingly integrated in systems deployed in full-fledged security operation centers~\cite{sans2025report,enisa2023threat,software2024revolutionizing,mink2023everybody,cybersecurity2025pulse,cloudflare2024threat,cloudflare2020intrusion,van2022deepcase} that protect the networks of modern organizations. Despite hundreds of works studying the security and robustness of ML-driven NIDS (ML-NIDS) in adversarial environments (see, e.g.,~\cite{apruzzese2021modeling,han2021evaluating,wang2022manda}), we assert that \textit{the real-world fidelity of the experimental approaches adopted in related research still remains an open issue.}

Recent seminal works have highlighted that real attackers operate in the ``problem space''~\cite{pierazzi2020intriguing} and must deal with complex ML systems~\cite{apruzzese2023real}, which can be outside the attacker's reach. And yet, numerous works persist in adopting an ML-centric view: for instance, subsequent works accepted to S\&P'24~\cite{diallo2024sabre} and EuroS\&P'24~\cite{bhusal2024pasa} used well-known ``gradient-based'' attacks~\cite{kurakin2016adversarial} to evade an ML-NIDS trained on the CICIDS17 benchmark dataset~\cite{sharafaldin2018toward}. 
Such adversarial perturbations are applied to the feature representation of a given input: despite substantially reducing the baseline accuracy (e.g., from 99\% to 16\%), these ``adversarial examples'' may not be physically realizable by a real-world attacker, who cannot violate strict domain constraints~\cite{sheatsley2021robustness}. Some works introduced perturbations in ``spaces'' that are closer to the attacker, such as manipulating pre-collected network packets~\cite{han2021evaluating,apruzzese2024adversarial}. Yet, even these perturbations are applied in the real problem space: to reliably change the packets analysed by an ML-NIDS, an attacker requires access to a highly privileged device (e.g., the router, or the NIDS itself~\cite{apruzzese2021modeling}). Is this possible? Without a doubt. But is this what an attacker would, or can, do in practice? {\footnotesize (This is an open question to the reader)}

\textbf{\textsc{Contributions}.} We posit that real attackers seeking to evade an ML-NIDS would first attempt to do so by issuing slightly different commands on the host they can directly control---and not by manipulating datapoints collected by other devices. \textit{We define such a tactic as a ``host-space perturbation''} (§\ref{sec:hsp}). As we will show with a systematic literature review across 316 papers (§\ref{sec:slr}), this specific class of adversarial perturbations has been overlooked by prior work. So, we took a first step (§\ref{sec:demo}) and assessed what happens if an attacker, instead of launching \command{patator persistent=1}{\small} in an attempt to bruteforce an ssh server~\cite{patator}, uses \command{patator persistent=0}{\small} instead: an ML-NIDS having perfect detection rate against the former command cannot detect a single NetFlow of the latter (we tested this both on benchmark datasets and in a real-world network). We thoroughly examine this use case (§\ref{sec:casestudy}), explaining the reasons behind these results, and also highlighting the pitfalls of replicating host-space perturbations by manipulating pre-collected data (i.e., the de facto way of applying adversarial perturbations). 
We also expand our assessment (in the Appendix~\ref{app:additional}) by considering additional attack types, affected ML models, and network environments.
Based on our findings, we therefore advocate for future work to revisit the approaches used for security assessments of ML-NIDS and provide recommendations on how to replicate host-space perturbations in a research setting (§\ref{sec:recommendations}).
We release all our experimental resources (for which we used open-source tools), including the data of our real-world evaluation, in our repository~\cite{repository}. 

\section{Background and Motivation}
\label{sec:background}

\noindent
We summarize the field of ML-NIDS (§\ref{ssec:nids}) and of adversarial ML~(§\ref{ssec:advml}) before making our problem statement~(§\ref{ssec:shortcomings}).

\subsection{Primer on ML-driven NIDS}
\label{ssec:nids}

\noindent
Research on Intrusion Detection has spanned almost forty years~\cite{denning1987intrusion}, and intrusion-detection systems (IDS) have been studied by thousands of publications~\cite{debar1999towards}. IDS are designed to \textit{detect intrusions}: in other words, they implicitly assume that an ``intrusion'' has taken place, and hence seek to warn users (e.g., by raising ``alerts'') as early as possible to contain the damage that an attacker may cause to the targeted system~\cite{lee2000framework}.

Over the years, IDS have been linked to a variety of terms covered in a plethora of taxonomies (e.g.,~\cite{debar2000revised,vasilomanolakis2015taxonomy}) and surveys~\cite{aleesa2020review,liao2013intrusion,khraisat2019survey,tsai2009intrusion}. Here, we consider a specific class of IDS: those using machine-learning (ML) techniques applied to network-related data. By ``ML techniques'', we intend automated data-driven methods whose detection capabilities are determined by a training phase, during which a given ML model (e.g., a decision tree, or a neural network) learns to make decisions by observing (typically ``large'') previously collected datasets. By ``network-related data'', we intend data pertaining to the communications occurring within a given network---such as, e.g., network traffic included in a packet-capture (PCAP) trace~\cite{fusco2012pcapindex}, or network flows (NetFlows) providing high-level summaries of the communications between two endpoints~\cite{vormayr2020my,barradas2021flowlens}. Such endpoints not necessarily involve different and/or physical devices (e.g., they can entail virtual machines, or even communications sent from, and directed to, the same host). We therefore denote our considered class of IDS as ML-based Network-IDS, or ML-NIDS (note that our focus overlaps with that of a recent SoK~\cite{apruzzese2023sok}).

\begin{figure}[!t]
\centering
\includegraphics[width=0.45\textwidth]{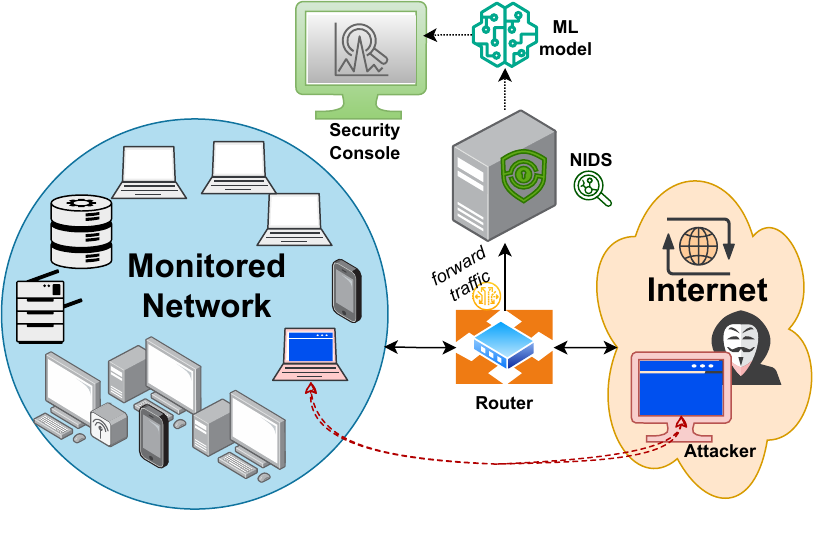}
\vspace{-5mm}
\caption{
{\small \textbf{Typical NIDS scenario.} \textmd{N.b.: the attacker \textit{cannot} control the router or the NIDS (otherwise, it wouldn't be surprising if the NIDS is bypassed.}}
}
\label{fig:nids}
\vspace{-2mm}
\end{figure}

We provide a visualization of an exemplary ML-NIDS deployment scenario in Fig.~\ref{fig:nids}. At a high level, the NIDS receives the network-related data pertaining to a monitored network. Specifically, such data is forwarded to the NIDS by a dedicated network appliance (e.g., a router). The NIDS then analyses such data by sending it to any given ML model---which can be seen as a classifier that ``detects'' if the data contains traces of an ``intrusion'' that must be escalated to a security operator, who must then take action. ML-NIDS have been widely studied in academic literature, and prior work showed that, e.g., ML-NIDS can detect malware~\cite{piskozub2021malphase,fu2021realtime}, botnet~\cite{garcia2014survey,mirsky2018kitsune}, DDoS~\cite{aktar2023towards,doshi2018machine}, reconnaissance~\cite{ge2019deep,viet2018using}, lateral-movement~\cite{rabbani2024graph,bai2019machine}, or bruteforcing~\cite{najafabadi2014machine,otoom2023deep} attacks. Indeed, even in the real world, ML methods are commonly deployed in security-operation centers, or by world-renowned networking companies~\cite{cloudflare2020intrusion} to detect malicious activities~\cite{sans2025report}.

\subsection{Adversarial Perturbations vs ML Models}
\label{ssec:advml}
\noindent
The growth of ML led to many studies scrutinising the security of these techniques when deployed in adversarial settings, giving rise to the field of ``adversarial machine learning''~\cite{huang2011adversarial}.

Thousands of papers have highlighted that ML models are vulnerable to ``adversarial perturbations'', i.e., tiny manipulations that adversely affect an ML model's performance~\cite{biggio2018wild}. 
There are various ways to convey such ``adversarial ML attacks'' (e.g.,~\cite{cina2024machine,biggio2013evasion,apruzzese2023real,papernot2018sok}). 
For instance, attackers can attempt to poison the ML model by tampering with its training phase~\cite{cina2024machine}; or attempt to evade its detection during its inference phase~\cite{apruzzese2023real}. 
Unfortunately, adversarial ML attacks can also target ML models designed for NIDS~\cite{apruzzese2021modeling}.

Assessment of adversarial ML attacks requires, from a research viewpoint, determining where (i.e., in which ``space'') the perturbation is applied~\cite{pierazzi2020intriguing}. Historically, adversarial perturbations had been applied in the \textit{feature space}, i.e., by directly changing the feature vector provided as input to the ML model.\footnote{E.g., changing the values of a ``malicious'' NetFlow to induce the targeted ML model to classify it as ``benign'', thereby evading an ML-NIDS~\cite{apruzzese2018evading}.} However, as pointed out by Pierazzi et al.~\cite{pierazzi2020intriguing}, \textit{real attackers operate in the ``problem'' space}---which not necessarily overlaps with the feature space~\cite{spacephish2022}. Indeed, as noted by more recent works~\cite{apruzzese2023real}, real-world attackers interact with ML systems---and not ML models. 

Such an observation highlights a crucial issue: in a lab setting, manipulations in the feature space are meant to approximate the operations that an attacker would perform in the real world; however, careless manipulation of the feature space risks creating ``adversarial examples'' that may not be physically realizable and/or which violate domain constraints~\cite{sheatsley2021robustness}.\footnote{For instance, manipulating the NetFlows may lead to numbers which would represent ``impossible-to-generate'' network traffic~\cite{sheatsley2023space}.}

\subsection{Shortcomings of Adversarial Perturbations}
\label{ssec:shortcomings}

\noindent
It can be said that the paper by Pierazzi et al.~\cite{pierazzi2020intriguing} questioned the real-world validity of the results portrayed by previous research, since feature-space perturbations were the norm\footnote{Most papers on adversarial ML focus on image recognition~\cite{apruzzese2023real,suya2024sok}, where ``feature-space'' perturbations entails manipulating pixels in an image~\cite{su2019one}.} in the adversarial ML domain---including the specific context of ML-NIDS~\cite{apruzzese2021modeling}. 

As a consequence, some works (e.g.,~\cite{wang2022manda,han2021evaluating}) began to explore scenarios wherein the perturbations were not applied to the input features, but in other spaces---allegedly being a better representation of a real attacker's operations. 
For instance, Han et al.~\cite{han2021evaluating} crafted ``traffic-space perturbations'' by manipulating a PCAP trace, adding bytes to the packets' payload: such changes would propagate to the feature space, thereby influencing the ML-NIDS; whereas Wang et al.~\cite{wang2022manda} claimed to generate ``problem-space adversarial examples'' by mutating network packets so that the corresponding feature representation (i.e., a NetFlow) was misclassified by the ML-NIDS. Despite the proven effectiveness of such perturbations (which led to correct feature vectors denoting malicious samples that bypassed the targeted ML-NIDS), we argue that \textit{such strategies may not resemble a real attacker's actions}.

Indeed, prior work is limited to applying perturbations by manipulating pre-collected datapoints---such as adding junk bytes to a packet in a PCAP trace. However, from an attacker's viewpoint, \textit{precisely} adding that exact amount of junk bytes to the specific packet (and just that one!) sent by their controlled host may be unfeasible (especially if the payload is encrypted); besides, the packets ``seen'' by the router (and forwarded to the NIDS) may not correspond to those sent by the attacker-controlled hosts.\footnote{For instance, consider the attack proposed in~\cite{sadeghzadeh2021adversarial}, described as follows: ``a dummy packet is injected into a specified location \scmath{k\in[0,n-1]} among the first \scmath{n} packets of a flow and an [universal adversarial perturbation] is injected into it.'' An attacker can certainly do so \textit{in theory}---but how, \textit{in practice}?} Put simply, to bypass an ML-NIDS, we argue that a real-world attacker may opt for more straightforward perturbation-related strategies---such as issuing different commands to their controlled hosts. We define such a class of adversarial ML attacks as ``host-space perturbations'' since they are physically applied to a (attacker-controlled) host.

\vspace{1mm}

{\setstretch{0.9}
\textbox{
\textsc{\textbf{Research Goal.}} We seek to: justify that ``host-space perturbations'' are a realistic (and, so far, overlooked) way to attempt evasion of ML-NIDS; and examine some of their effects.}}

\vspace{1mm}
\textbf{Focus of the paper.} In this work, we consider NIDS and, specifically, those focusing on network traffic analysis. Orthogonal application domains, which are outside our scope, include: IDS for Controller Area Networks, since they use different datatypes (e.g.,~\cite{cerracchio2024investigating,shahriar2023cantropy}); or for industrial-control/cyber-physical systems, since they mostly rely on sensor data (e.g.,~\cite{luo2021deep,erba2020constrained,erba2023practical}); as well as anti-malware engines~\cite{alasmary2019analyzing}, or IDS analysing provenance data (e.g.~\cite{alsaheel2021atlas,jia2024magic,cheng2024kairos}). 
\begin{figure*}[!t]
    \centering
    \includegraphics[width=1.75\columnwidth]{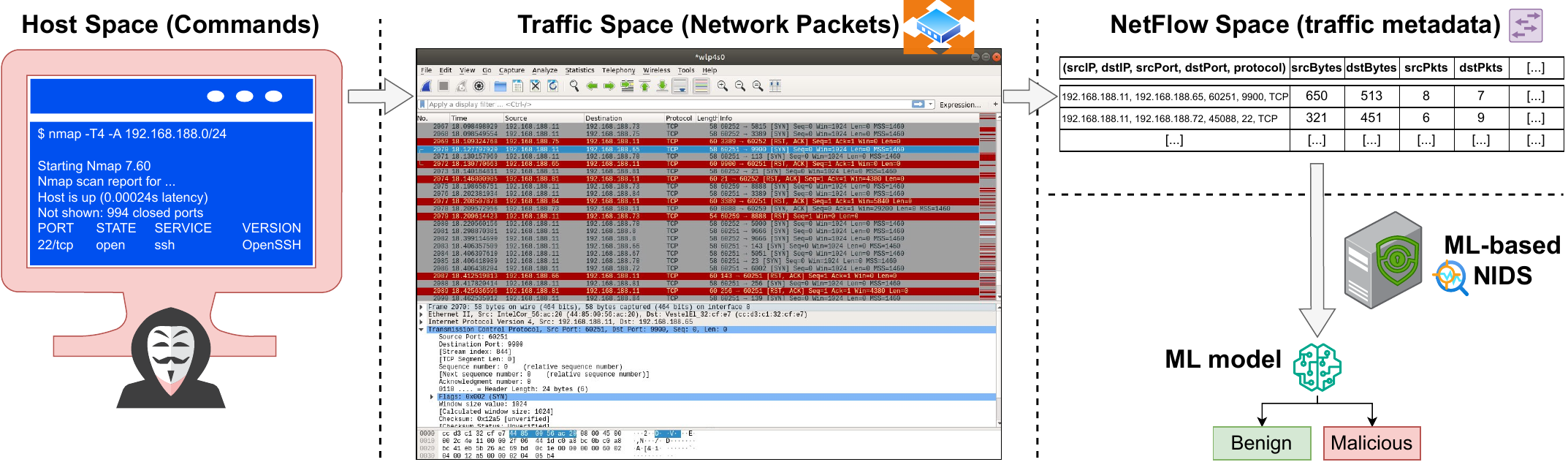}
    \vspace{-2mm}
    \caption{\textbf{From attackers' actions to ML inputs.} \textmd{[Left] The attacker launches \command{nmap}{\scriptsize} on their controlled host. [Middle] This leads to the creation of multiple network packets that are captured by some dedicated network appliance (e.g., a router). [Right] Then, NetFlows are extracted from the PCAP trace, which are sent to (and analysed by) the ML-NIDS. A realistic attacker has no access to the ``traffic space'' (i.e., middle panel) and ``NetFlow space'' (i.e., right panels): the attacker can only operate at the host level (i.e., left panel). Therefore, directly manipulating the PCAP trace (captured by the router), the NetFlows (extracted from the PCAP trace), or their feature representation (analysed by the ML model) would violate the assumptions of the threat model.}}
    \vspace{-2mm}
    \label{fig:spaces}
\end{figure*}

\section{Host-space Perturbations}
\label{sec:hsp}

\noindent
We present our primary contribution: the conceptualization of ``host-space'' adversarial perturbations~(HsP). HsP can be seen as an \textit{alternative way} to attack (or assess the robustness of) ML-NIDS. We first introduce HsP with a reflective analysis (§\ref{ssec:intuition}), then define HsP~(§\ref{ssec:threat}) and provide some security considerations on HsP~(§\ref{ssec:security}).

\subsection{Generic Intuition}
\label{ssec:intuition}

\noindent
In security assessments, the researcher should aim at reproducing the workflow that a hypothetical attacker would follow in the real world~\cite{apruzzese2023real}.  
\ul{In the context of adversarial perturbations, a crucial question that must be answered is: ``What is the attacker's problem space?''}\footnote{We provide in Appendix~\ref{app:perturbations} a discussion on how to tackle this question in adversarial ML domains orthogonal to that of NIDS (e.g., phishing, malware, or computer vision).} 
We answer this question by putting ourselves in the attacker's shoes: through \textit{inductive reasoning}~\cite{carroll2012realizing}, we demonstrate that, in the real world, (most) attackers would (and, likely, can) only operate on the ``host space'', i.e., on the hosts they control.

\textbf{Context.} Let us assume that an attacker seeks to carry out some malicious operation within a given network. Without loss of generality, such an objective can entail any type of (malicious) network communication: external-to-internal (e.g., a remote DDoS attack); internal-to-external (e.g., data exfiltration); or internal-to-internal (e.g., reconnaissance, lateral movement). The network is protected by some NIDS, which can be deployed anywhere---but, for simplicity, we assume that the NIDS is deployed at the network's border. The NIDS integrates some ML model that analyses network-related data of any type (e.g., packets payload~\cite{song2020software}, or NetFlows~\cite{wang2022enidrift}): without loss of generality, we assume that the NIDS analyses NetFlows, since it allows broad coverage (see e.g.,~\cite{bajaber2024p4control,sendner2024mirageflow} or recent surveys~\cite{apruzzese2023sok, verkerken2025rustiflow}). In this context, the attacker performs ``actions'' on their controlled host (external or internal); such actions lead to the generation of network packets, which are preprocessed into NetFlows, and then fed to an ML model within the NIDS that determines whether the NetFlows are benign or malicious. Hence, in this context, \ul{the problem space wherein the attacker operates is defined by the \textit{actions that the attacker can perform on their controlled host}}. 

\textbf{Exemplary use-case.} Assume the attacker, who has obtained remote access to an internal host, wants to carry out some reconnaissance: the attacker uses the \command{nmap}{\small} command, specifying any parameter (e.g., \command{-T4 -A}{\small}). Such actions generate some packets and NetFlows, whose values depend on the parameters specified by the attacker. This workflow is illustrated in Fig.~\ref{fig:spaces}. We observe that the packets are captured by the router/switch (to which the attacker has no access), which forwards them to the NetFlow extractor (to which the attacker has no access), which sends the NetFlows to the ML-NIDS (to which the attacker has no access). Hence, the attacker has some control over the packets sent by their controlled host, but once these packets ``leave'' this host, our envisioned attacker cannot modify them in any way. Simply put, such an attacker cannot manipulate the data captured by a router/switch, nor the resulting NetFlows, nor the individual feature-vectors analysed by the ML-NIDS---\textit{all of which being ways in which prior work applied adversarial perturbations} (we will justify this assertion in §\ref{sec:slr}). 
If the attacker suspects that the ML-NIDS is trained to detect reconnaissance attempts issued with \command{nmap -T4 -A}{\small}, the attacker may take another action, e.g., using \command{-T0}{\small} (instead of \command{-T4}{\small}): doing so would lead to a slower scan, meaning that different packets and, hence, different NetFlows will be generated by their host.
Therefore, the perturbations that can be crafted by real-world attackers pertain to the actions (in terms of commands launched) performed on their \textit{controlled hosts}---i.e., the ``problem space.''

\vspace{-1mm}

\begin{cooltextboxgreen}
\textsc{\textbf{Takeaway.}} Real attackers perform \textit{actions} leading to changes in the network data generated by their controlled hosts. From an adversarial ML viewpoint, this can be simulated through ``host-space perturbations'', which require operating at the host level---and not on the data that, despite having been generated by such a host, was captured by an appliance \textit{outside the attacker's control}.
\end{cooltextboxgreen}

\subsection{Definition and High-level Threat Model}
\label{ssec:threat}

\noindent
We are not aware of any existing definition of ``host-space perturbation'' (which are conceptually different from the ``problem-space transformations'' of~\cite{pierazzi2020intriguing}; we discuss this in~§\ref{ssec:security}). 
Hence, we provide our own. Broadly, we define a host-space perturbation as ``\textit{any operation that allows an attacker to achieve the same goal by executing commands or actions on the hosts under their control (as defined by the specified threat model)}''. Here, a ``goal'' refers to the successful execution of an attack. A more formal definition is as follows.

\vspace{1mm}

{\setstretch{0.99}
\textbox{
Assume an attacker who controls a (set of) hosts \smacal{H} and aims to achieve a given goal \smacal{G} through a series of operations \smacal{O} executed on those hosts. A host-space perturbation is defined as \textit{any alternative set of operations} \smacal{O'}\smamath{\neq}\smacal{O} that allows the attacker to achieve the same goal \smacal{G} by operating on the same set of hosts \smacal{H}. It is implicitly assumed that both \smacal{O} (and \smacal{O'}) generate network traffic analysed by the ML model that the attacker aims to evade via the host-space perturbation---though evasion is not guaranteed.}}

\vspace{1mm}

We observe that, according to our definition, an HsP does not identify a precise ``perturbation''.\footnote{HsP are \textit{complementary} to traditional adversarial paradigms in the ML-NIDS context.} Rather, the HsP is defined as an alternative way to reach the same goal. Note, however, that such a ``goal'' can be defined either broadly or more narrowly. For instance, a broad goal can be ``carrying out a port scan''; a more narrow one is ``carrying out a port scan in a low-and-slow manner''~\cite{darktrace2018lowandslow}: and a very specific one can be ``carrying out a port scan in a low-and-slow manner by using \command{nmap}{\small}''. Clearly, a broad goal implies a great number of ``alternative ways'' in which such a goal can be achieved. 

We can hence abstract our generic definition and provide a formal threat model for HsP. To align such a threat model with established works, we follow the guidelines of~\cite{apruzzese2023real}, who recommend defining a threat model by adopting a system-wide view.

\begin{cooltextbox}
\textsc{\textbf{Threat Model.}} Given an attacker having a goal \smacal{G}, which they pursue by using their knowledge \smacal{K} and capabilities \smacal{C} on the targeted system to devise a certain strategy \smacal{S}. To realise a valid HsP, the following must be met:

\begin{itemize}[leftmargin=*]
    \item \textbf{Goal.} The overarching goal is the same as in the original threat model (i.e., \smacal{G}), but it must also include evasion of an (ML-based) NIDS.\footnote{Otherwise, the attacker would not even attempt any evasion.} (By ``evasion'', we intend a misclassification of malicious samples at test/inference-time~\cite{apruzzese2023real}.)
    \item \textbf{Knowledge.} The knowledge is identical to that of the original threat model (i.e., \smacal{K}), but it must be assumed that the attacker expects that the ML-NIDS would detect the strategy \smacal{S} the attacker would follow if they did not apply any HsP.\footnote{Otherwise, why would the attacker even attempt using an HsP?}
    \item \textbf{Capabilities.} The capabilities must not change w.r.t. \smacal{C}: the attacker cannot be assumed to, e.g., control more hosts, or have more privileges on their already controlled hosts.
    \item \textbf{Strategy.} The strategy must change w.r.t. \smacal{S}, and is still dictated by the assumed \smacal{K} and \smacal{C}, and must only include operations that involve the attacker's controlled hosts.\footnote{E.g., an attacker cannot ``bribe'' an employee to obtain a password; or physically go to a host they wish to portscan in order to check which ports are open locally.}
\end{itemize}
\end{cooltextbox}

\subsection{Security Considerations}
\label{ssec:security}

\noindent
Let us analyse our proposed HsP from a security standpoint (we provide additional ``critical'' remarks in §\ref{ssec:critical}).

First, our ``requirements'' on the \textit{goal} and \textit{knowledge} are also implicit in any threat model entailing attacks implemented via adversarial perturbations. For instance, no attacker would, e.g., attempt to introduce a gradient-based perturbation if they did not expect the existence of a detector that would flag their ``non-adversarial-examples.'' Similarly, it is implicit that, when choosing a ``new'' strategy, attackers would favor simple ones (e.g., guessing-based strategies~\cite{apruzzese2023real}), which not necessarily are the most optimal ones.

Second, ``host-space perturbations'' satisfy the required properties of ``problem-space adversarial ML attacks'' provided by Pierazzi et al.~\cite{pierazzi2020intriguing}: this is because the perturbation is, by definition, applied in the attacker's problem space. However, and crucially, HsP are not ``problem-space \textit{transformations}''. Indeed, realising an HsP implies \textit{physically} executing a different set of operations (\smacal{O'}\smamath{\neq}\smacal{O}). There is no ``transformation'' in an HsP: the principle is not to ``create an adversarial example by applying problem-space transformations to a given (non-adversarial) example''; rather, the point is ``creating new datapoints entirely \textit{in the problem space}''. And indeed, this is why we emphasized that ``evasion is not guaranteed''. Put differently, an HsP should be seen as a complementary way through which researchers should investigate the security/robustness of (ML-driven) NIDS \textit{by directly operating in the attacker's problem space} (i.e., on the hosts controlled according to \smacal{C}).

Finally, we further illustrate some intriguing properties of HsP with an example. Let us consider the simple use case (from §\ref{ssec:intuition}) of an attacker whose goal (\smacal{G}) is to ``port-scan'' some device. The attacker does so by issuing \command{nmap -T0}{\small} (i.e., \smacal{O}) on their controlled host (\smacal{H}). The attacker can also achieve the same goal (\smacal{G}) from the same host (\smacal{H}) by issuing \command{nmap -T4}{\small} (\smacal{O'}). We use this example to make four important considerations on our host-space perturbations.
\begin{itemize}[leftmargin=*]
    \item \textit{Certain host-space perturbations are easier-to-apply than others}. E.g., changing a parameter from \command{T0}{\small} to \command{T4}{\small} is a simple and valid host-space perturbation. The attacker could also manually probe each port using \command{netcat}{\small}; despite being more labour-intensive, it is a valid host-space perturbation (provided, of course, that \command{netcat}{\small} is available on the attacker's controlled hosts).
    
    \item Some host-space perturbations may cause substantial changes in the ``feature space'', potentially leading to \textit{out-of-distribution} (OOD) samples that are misclassified. However, \ul{this is not a concern for the real attacker}~\cite{apruzzese2023real}, whose objective is evading the ML-NIDS while achieving their overarching goal (e.g., a port scan in the previous example). HsP should not be seen as ways to induce ``minimal'' feature-space changes (the focus of most adversarial-ML papers, such as~\cite{pierazzi2020intriguing}).
    
    \item An \textit{invalid host-space perturbation} involves an activity that is not admitted in the threat model---for instance, executing a command that is not assumed to be available on the attacker-controlled host (e.g., running \command{sudo}{\small} without having root privileges), since this violates the envisioned capabilities (i.e., \smacal{C}).
    
    \item From a researcher's viewpoint, \textit{it is unwise to simulate the effects of host-space perturbations via manipulations of pre-collected data}. For instance, simulating the change from \command{-T4}{\small} to \command{-T0}{\small} by manipulating the NetFlows (i.e., a ``feature-space'' perturbation, as done in~\cite{apruzzese2018evading,schneider2021evaluating}) can lead to inconsistent adversarial examples~\cite{sheatsley2021robustness} (we are not aware of a one-way function that goes from ``command'' to ``NetFlow''); whereas manipulating the PCAP trace (as done, e.g., by~\cite{han2021evaluating}) is also unreliable because networks are ``unpredictable'' ecosystems~\cite{apruzzese2023sok}: for instance, going from \command{-T4}{\small} to \command{-T0}{\small} may lead to some packets being lost or retransmitted. 
\end{itemize}
In our assessment (in §\ref{sec:demo}) we show ``easy-to-apply'' HsP, as well as of (potentially) ``invalid'' HsP, and we also show what happens when attempting to approximate HsP (in §\ref{ssec:pitfall}).
\section{Systematic Literature Review (RQ1)}
\label{sec:slr}

\noindent
To scrutinize the novelty of HsP in the adversarial ML context applied to the NIDS domain, we ask ourselves our first research question (RQ1): 
``{\color{violet}Has prior work on the robustness of ML-NIDS considered perturbations involving launching different commands on the attacker-controlled host (i.e., HsP)?}'' We answer RQ1 with a systematic literature review (SLR).

\vspace{-0mm}

\subsection{Research Method}
\label{ssec:slr_method}

\noindent
To answer RQ1, we must fulfill two objectives: {\small \textit{(i)}}~identify papers investigating the ``robustness of ML-NIDS,'' and
{\small \textit{(ii)}}~inspect their content to determine if the assessment entails HsP.

To tackle the first objective, we adopt an approach similar to recent works~\cite{arp2022and, apruzzese2023sok}, which focus on identifying papers from a select subset that is likely to include literature relevant for RQ1. We consider all the works that are ``linked'' to six pivotal papers:~\cite{wang2022manda, han2021evaluating, apruzzese2023sok, apruzzese2023real, flood2024bad, apruzzese2021modeling}. These works all deal with the subject of ML-NIDS and either {\small \textit{(a)}}~have been published in top-tier venues~\cite{flood2024bad,apruzzese2023sok,apruzzese2023real,wang2022manda}; or {\small \textit{(b)}}~cover a large number of previously-published papers on, among others, ``robustness of ML-NIDS''~\cite{apruzzese2023sok,apruzzese2021modeling, flood2024bad}; or {\small \textit{(c)}}~carry out their evaluations by using perturbation techniques that are methodologically close to HsP~\cite{wang2022manda,han2021deepaid}. Hence, these works are a valid starting point to select the literature used to answer RQ1.

To tackle the second objective, we adopt a mix of automated keyword-based filtering, and manual qualitative analysis. The latter is done by employing the \textit{dual-reviewer system} (as also done in~\cite{arp2022and,apruzzese2023real,schroer2025sok}): two researchers (having [6--9] years of experience in ML-NIDS) inspected the papers and, in case of doubt, discussed their findings to reach a consensus. Altogether, these operations have been done twice: once in August 2024, and a second time in September 2025 (to update our findings with more recent works).

\subsection{Systematic Workflow Description}
\label{ssec:slr_workflow}
\noindent
We describe our SLR by following established PRISMA guidelines~\cite{page2021prisma}, providing the results after each major step.

\textbf{Paper collection.} First, we consider: the 38 works analysed by~\cite{flood2024bad}; the 46 papers reviewed in~\cite{apruzzese2023sok}; and the 88 papers in~\cite{apruzzese2023real}. We do so because all these works scrutinised literature related to ML-NIDS (for~\cite{apruzzese2023sok, flood2024bad}) or ML security (for~\cite{apruzzese2023real}) published in top-tier venues within 2017--2023. Then, we consider~\cite{han2021evaluating,apruzzese2021modeling,wang2022manda}, and apply the snowball method~\cite{wohlin2014guidelines}, collecting all papers that are cited by (according to Google Scholar) either of these works~\cite{han2021evaluating,apruzzese2021modeling,wang2022manda} and which were published until the end of 2024. We do so because these three works~\cite{han2021evaluating,wang2022manda,apruzzese2021modeling} are well-known in the ML-NIDS domain and consider ``problem space'' adversarial ML attacks that resemble real-world scenarios. After removing duplicates, we obtain a set of 316 papers (note: we consider all works returned by Google Scholar, including gray literature, as also done in~\cite{schroer2025sok}).

\textbf{Automated Filtering.} We review these 316 papers and determine whether there is an evaluation of ``adversarial ML attacks'' against ML-NIDS. We do so by means of a keyword search with the terms ``adversarial perturbation / attack / example''. We filter the papers mentioning any of the keywords for at least three times in their full text (many papers simply mention similar terms for future work, e.g.~\cite{barradas2021flowlens}), increasing the likelihood that the paper is indeed about adversarial ML; this gives us 201 papers.

\textbf{Manual Analysis.} We analyse the remaining 201 papers that carry out an ``adversarial'' assessment of ML-NIDS (excluding, e.g., literature reviews~\cite{apruzzese2021modeling}). We scrutinise ``where'' the perturbation is applied: this is done with a qualitative analysis focused on inspecting the experimental methodology adopted in the paper, gauging whether the perturbations stem from attackers issuing different commands on their controlled hosts (i.e., host-space perturbations); we may also check the source code (if provided).\footnote{We also scrutinised if the experiments entailed an original real-world network-data collection step (without, e.g., exclusively relying on benchmark datasets). We found only 12 works (out of 201, i.e., 5.9\%) doing so:~\cite{silva2024gonogo,ndichu2024adversarial,sanchez2024adversarial,
barradas2021flowlens, okada2024xai, catillo2024towards, yan2023automatic,
salman2022mutated, jan2020throwing, nasr2021defeating, aiken2019investigating,
abdelaty2021gadot}}

\subsection{Findings and Analysis}
\label{ssec:slr_findings}

\noindent
Across the 201 papers we manually analysed, the perturbations are predominantly crafted by directly modifying the data---potentially before preprocessing (e.g.,~\cite{abdelaty2021gadot}), or via feature-space manipulations (some of which preserving domain constraints~\cite{severi2023poisoning}). 
We mention two noteworthy approaches:
\begin{itemize}[leftmargin=*]
    \item Catillo et al.~\cite{catillo2024towards}, explore ``problem-space perturbations'' by introducing delays through modifications to the attacker's host network configuration using Linux's Traffic Control utility~\cite{tc}. However, this approach requires the attacker to have admin power on the host, and has the drawback of affecting \textit{all} traffic between this specific host and the target---including the ``benign'' communications. Under the assumption that the attacker has such a power, this could qualify as a valid HsP. Yet, we could not find any ``threat model'' or detailed description of the capabilities of the envisioned attacker in~\cite{catillo2024towards}.

    \item The adversarial perturbations crafted in~\cite{apruzzese2024adversarial} are applied to the PCAP trace by using the Python library scapy~\cite{scapy} to add junk bytes to certain network packets. Such an approach is, by definition, \textit{not} an HsP. Although the perturbations are not in the ``feature space'' (since the targeted ML model analyses NetFlows, whereas the perturbations are applied to network packets) it is unclear whether the corresponding ``manipulated'' network packets would be received by the involved host exactly in the same order as that reported in the PCAP trace: indeed, appending junk bytes means that the packets are ``larger'', which leads to more bandwidth being used, and which translates to potential (and ultimately unpredictable) differences in the way the traffic is distributed across the whole network.\footnote{We reached out to the authors of~\cite{apruzzese2024adversarial}, who confirmed this fact.} However, we stress that this approach does preserve domain constraints.    
\end{itemize}
Even papers published in 2025 (outside our SLR) apply direct data manipulations, such as~\cite{brockmann2025now} which acknowledge ``we cannot guarantee that all the adversarial examples can be produced in a real world setting.'' Hence, altogether, these findings echo a statement made in a recent work: ``Problem-space Evasion Adversarial Attacks are Already Extremely Hard for an Attacker''~\cite{elshehaby2026novel}.

Nonetheless, our analysis elucidates a blind spot in ML-NIDS research. To the best of our knowledge, it is unknown how resilient current ML-NIDS are against HsP. We hypothesize that \textit{this gap stems from the practical challenges of studying HsP in a research setting}: it requires access to an operational networked host, executing ``adversarial commands,'' capturing the corresponding traffic, generating the feature representation, and studying the effects on the ML-NIDS. These steps are non-trivial, especially since most prior work relies on public datasets~\cite{flood2024bad} collected in networks (and, hence, hosts) not accessible by downstream researchers. 

\begin{cooltextboxgreen}
\textsc{\textbf{Answer to RQ1:}} We could not find any paper that allows to answer positively to RQ1. Hence, according to our SLR (and to our knowledge), no prior work has attempted to ``adversarially evade'' an ML-NIDS by operating (e.g., launching different commands) on the attacker-controlled host---i.e., our proposed HsP.
\end{cooltextboxgreen}

\section{Demonstration~of~Host-space~Perturbations}
\label{sec:demo}

\noindent
Given the negative answer to RQ1, we wonder (RQ2): ``{\color{violet}what are some effects that HsP can have on existing ML-NIDS?}'' To tackle RQ2, we carry out a proof-of-concept evaluation of our proposed HsP.\footnote{Across all our experiments, we do not ``data snoop'~\cite{arp2022and} to inflate the performance of our ML models---we have no incentive to do so. We maintain a clear separation between training and test data, never evaluating on samples seen during training. Moreover, since our data is collected over a short timeframe, a temporal split is unnecessary. Prior empirical studies on similar testbeds have shown that it yields no significant difference/benefit~\cite{apruzzese2023sok}. Additional data-collection and pre-processing details are provided in the Appendix~\ref{app:data}, whereas complete details of our experiments (e.g., hyperparameters), including the underlying source code, are provided in our repository~\cite{repository}.} Such a demonstration has a twofold goal: {\small \textit{(i)}}~show how HsP can be practically leveraged (by real-world attackers) and realised (by researchers), and {\small \textit{(ii)}}~examine their effects on well-known ML-based NIDS (i.e., the crux of RQ2).

\subsection{Experiments on Benchmark Datasets}
\label{ssec:demo_benchmark}
\noindent
We begin our demonstration by focusing on ML-NIDS trained and tested via the well-known CICIDS17 and CICIDS18 datasets~\cite{sharafaldin2018toward}. These datasets, which count over 3k citations on Google Scholar, are widely used to benchmark ML-based NIDS focused on NetFlows. Therefore, they are ideal for our assessment. Importantly: we use the \textit{fixed version} of these datasets (provided in~\cite{engelen2021troubleshooting} and~\cite{liu2022error}).

\textbf{Threat Model.} We envision an ML-NIDS that is trained on the data contained in either the CICIDS17 or CICIDS18 datasets. These datasets present similarities: the latter extends the former by including more ``benign'' data, as well as ``malicious'' data pertaining to a wider range of attacks. Both datasets have malicious datapoints related to the \command{patator}{\small} attack, which is a well-known ssh-bruteforcing tool~\cite{patator}. Specifically, these datasets include network-traffic data captured by launching \command{patator {-}{-}persistent=1}{\small}. For an exemplary demonstration, we assume an attacker who, after having compromised a host within the network monitored by the ML-NIDS, wants to ssh-bruteforce a given ssh server while bypassing such an ML-NIDS (i.e., \smacal{G}). In terms of capabilities (i.e., \smacal{C}), the attacker can launch arbitrary commands on their compromised host, but does not have root or physical access to it, and has no control on any other host in the network. Given that the attacker expects the ML-NIDS to be trained to detect \command{patator {-}{-}persistent=1}{\small} (i.e., \smacal{K}), the attacker launches \command{patator {-}{-}persistent=0}{\small} (i.e., the new strategy). Such a strategy entails a valid HsP, which only requires changing a single character of the command and does not require any additional privilege on the attacker-controlled host (meaning that \smacal{C} does not change).

\begin{table}[t]
    \caption{\textbf{Experiments on benchmark data.} \textmd{\footnotesize Results (\scmath{fpr}/\scmath{tpr} on benign/malicious NetFlows) of our models for each ``train set'' and ``test set'', for both the CICIDS17 and CICIDS18. Boldface denotes NetFlows generated artificially via~\cite{verkerken2026concap}. We report the std in our repository~\cite{repository}. We use ``P'' to denote the option ``{-}{-}persistent'' of \command{patator}{\scriptsize}.}}
    \vspace{-3mm}
    \label{tab:demo_benchmark}
    \centering
    \resizebox{0.98\columnwidth}{!}{
        \begin{tabular}{c?rrr|rrr?rrr}
        \toprule
        Dataset & \multicolumn{6}{c?}{CICIDS17} & \multicolumn{3}{c}{CICIDS18} \\ \hline
        Train Set & \multicolumn{3}{c|}{Benign + P=1} & \multicolumn{3}{c?}{Benign + all malicious} & \multicolumn{3}{c}{Benign + P=1} \\
        Test Set & {Benign} & {P=1} & \textbf{P=0} & Benign & (all) & \textbf{P=0} & {Benign} & {P=1} & \textbf{P=0} \\
        \midrule
        DT & $<$0.001 & 0.997 & \cellcolor[HTML]{FAD9D5} 0.224 & $<$0.001 & $>$0.999 & \cellcolor[HTML]{FAD9D5} 0.214 & 0.000 & 1.000 & \cellcolor[HTML]{FAD9D5} 0.000 \\
        RF & 0.000 & 0.998 & \cellcolor[HTML]{FAD9D5} 0.490 & $<$0.001 & $>$0.999 & \cellcolor[HTML]{FAD9D5} 0.073 & 0.000 & 1.000 & \cellcolor[HTML]{FAD9D5} 0.000 \\
        XGB & $<$0.001 & 0.998 & \cellcolor[HTML]{FAD9D5} 0.986 & $<$0.001 & $>$0.999 & \cellcolor[HTML]{FAD9D5} $<$0.001 & 0.000 & 1.000 & \cellcolor[HTML]{FAD9D5} 0.000 \\
        SVM & $<$0.001 & 0.993 & \cellcolor[HTML]{FAD9D5} 0.000  & $<$0.001 & 0.997 & \cellcolor[HTML]{FAD9D5} $<$0.001 & 0.000 & 1.000 & \cellcolor[HTML]{FAD9D5} 0.000 \\
        DNN & 0.000 & 0.998 & \cellcolor[HTML]{FAD9D5} 0.000 & $<$0.001 & $>$0.999 & \cellcolor[HTML]{FAD9D5} $<$0.001 & 0.000 & 1.000 & \cellcolor[HTML]{FAD9D5} 0.000 \\

        \bottomrule
        \end{tabular}
    }
    \vspace{-1mm}
\end{table}

\textbf{Baseline Setup.} We carry out three experiments---all conforming to our underlying threat model. First, we train five well-known and ML-based classifiers -- namely: decision tree (DT), random forest (RF), histogram-gradient boosting (XGB), support vector machine (SVM), and a deep neural network (DNN) -- on a training set containing 80\% of the benign data in CICIDS17, and on 80\% of the data pertaining to \command{patator {-}{-}persistent=1}{\small} in CICIDS17; then, we test such models on the remaining 20\% (benign, and of \command{patator {-}{-}persistent=1}{\small}) in CICIDS17. Next, we do the same, but for CICIDS18 (thereby obtaining five different models). Next, we train another set of five models, this time using 80\% of all malicious data in CICIDS17 (not just the datapoints of \command{patator {-}{-}persistent=1}{\small}) alongside 80\% of its benign data, and we test these models on the remaining 20\% datapoints (benign and malicious). We always repeat this process five times, averaging the results. Altogether, these tests reveal that our models exhibit high \smamath{tpr} (above 0.999) and low \smamath{fpr} (below 0.001), i.e., they align with state-of-the-art performance reported in~\cite{liu2022error,engelen2021troubleshooting}.

\textbf{Attack implementation and results.}
To test the HsP, \ul{we need to create new data from scratch}, conforming to \command{patator {-}{-}persistent=0}{\small}. This is because there is no datapoint in either CICIDS17 or CICIDS18 that captures such an attack. To this end, we use the open-source network-simulator tool proposed in~\cite{verkerken2026concap}, which allows users to configure an ad-hoc network environment (which we configured so as to resemble to network environment in CICIDS17 and CICIDS18) and automatically generate labeled data pertaining to attacks launched by a given ``attacker'' host against a given ``target'' host (we configured the ``attacker'' and ``target'' host so that they are identical to the machines used in CICIDS17/18 to perform the \command{patator {-}{-}persistent=1}{\small} attack, but of course we specified the command \command{patator {-}{-}persistent=0}{\small}). This way, we obtained a set of labeled NetFlows that exactly reproduces the behavior of our envisioned attacker. We then submit these NetFlows to the respective ML models, and test their performance. The results of all these experiments are reported in Table~\ref{tab:demo_benchmark}. We can see that the \smamath{tpr} of our models against the malicious NetFlows of \command{patator {-}{-}persistent=0}{\small} is substantially lower than that achieved against \command{patator {-}{-}persistent=1}{\small}. The only exception is the XGB trained only on the malicious NetFlows of \command{patator {-}{-}persistent=1}{\small} on CICIDS17, which obtains \smamath{tpr}=0.986 but which drops to $<$0.001 when trained on all malicious data of CICIDS17 (this is a clear case of when generalizability across a wide range of malicious attacks reduces the performance against specific attacks). On CICIDS17, the DT and RF are the only models whose \smamath{tpr} is not $<$0.001, but the resulting values indicate that these models would be bypassed by such an HsP. On CICIDS18, all models are defeated.

\subsection{Real-World Network Experiment}
\label{ssec:demo_real}
\noindent
The previous experiments were done on ``benchmarks'', and we had to resort on a dedicated tool to create datapoints that would enable testing an HsP. Here, we assess HsP \textit{on a real-world network}.

\textbf{Threat Model.} We envision a similar threat model as in the previous experiment. However, here, the targeted network is a smart-home network comprising 35+ active hosts, including laptops, desktops, gaming consoles, and various IoT appliances (e.g., light bulbs, smart switches). In this network (which is much larger and recent than that in CICIDS17), two devices are set up so as to enable SSH connections between them. The adversary is assumed to control one of these devices, and seeks to bruteforce his/her way onto the other device by means of the \command{patator}{\small} tool. The network is monitored by an ML-NIDS that is specifically trained to recognize either \command{patator {-}{-}persistent=0}{\small} (in which case, the attacker would use \command{patator {-}{-}persistent=1}{\small}) or \command{patator {-}{-}persistent=1}{\small} (in which case, the attacker would use \command{patator {-}{-}persistent=0}{\small}). Note that all such assumptions align with our overarching threat model (in §\ref{ssec:threat}) while also enabling alignment with the previous experiment (in §\ref{ssec:demo_benchmark}). 

\textbf{Baseline Setup.} We follow a similar protocol as in the previous experiment. First, we capture a large PCAP trace (of \smamath{\approx}5GB, representing 12h of network activities) from the monitored network: we checked this trace and confirmed that no malicious activities were present. Then, we used the device assumed to be controlled by the attacker (i.e., a laptop running Ubuntu 20.04) to launch \command{patator {-}{-}persistent=0}{\small} against the ``targeted'' ssh server; we repeated this process also for \command{patator {-}{-}persistent=1}{\small}. We captured the traffic, and generated the corresponding NetFlows. We manually labeled such NetFlows, ensuring that only those pertaining to \command{patator}{\small} were labeled as malicious. Then, we trained our five baseline models (DT, RF, XGB, SVM, DNN): first, we took 80\% of the benign data and 80\% of the data of \command{patator {-}{-}persistent=1}{\small}; then, we trained another set of five baseline models, this time on 80\% of the benign data and 80\% of the data of \command{patator {-}{-}persistent=0}{\small}. We then tested them on the remaining 20\%: the performance (both \smamath{tpr} and \smamath{fpr}) was always near-perfect. (As before, we repeated this procedure five times, averaging the results). To further confirm the effectiveness of our models, we even tested them on a larger (benign) PCAP trace of 15GB, confirming that they always yield near-zero \smamath{fpr}, indicating that they are ``good'' because they do not raise false positives.

\textbf{Attack results.} To gauge the effectiveness of the HsP, we tested the models trained on \command{patator {-}{-}persistent=0}{\small} against the NetFlows produced by \command{patator {-}{-}persistent=1}{\small}, and vice-versa. The results of this entire experiment are reported in Table~\ref{tab:demo_real}. We see that our HsP are very effective: out of the models trained on \command{patator {-}{-}persistent=1}{\small}, only one does not have \smamath{tpr}=0: the SVM has \smamath{tpr}=0.49, which is still a substantial drop from the perfect \smamath{tpr} achieved on \command{patator {-}{-}persistent=1}{\small}. For the models trained on \command{patator {-}{-}persistent=0}{\small}, two (the SVM and the DNN) are not affected, but the remaining three are completely defeated. Note that these results validate those we obtained on the experiments on benchmark datasets: in all cases, changing a single character is enough to bypass most models.

\begin{table}[t]
    \caption{\textbf{Experiments on real-world data.} \textmd{\footnotesize Results (\scmath{fpr} on benign and \scmath{tpr} on attack NetFlows) of our models for each ``train set'' and ``test set'' (averaged over 5 trials; we provide the std.dev. in our repository~\cite{repository}).}}
    \label{tab:demo_real}
    \vspace{-3mm}
    \centering
    \resizebox{0.9\columnwidth}{!}{
        \begin{tabular}{l?rrr|rrr}
        
        \toprule
        Train Set & \multicolumn{3}{c|}{Benign + \command{P}{\small}=1} & \multicolumn{3}{c}{Benign + \command{P}{\small}=0} \\ \hline
        
        Test Set & Benign & \command{P}{\small}=1 &  \command{P}{\small}=0 &  Benign &  \command{P}{\small}=0 & { \command{P}{\small}=1} \\
        \midrule
        DT & $<$0.001 & 1.000 & \cellcolor[HTML]{FAD9D5} 0.000 & $<$0.001 & 1.000 & \cellcolor[HTML]{FAD9D5} 0.000 \\
        RF & 0.000 & 1.000 & \cellcolor[HTML]{FAD9D5} 0.000 & 0.000 & 1.000 & \cellcolor[HTML]{FAD9D5} 0.000  \\
        XGB & 0.000 & $>$0.999 & \cellcolor[HTML]{FAD9D5} 0.000 & $<$0.001 & 1.000 & \cellcolor[HTML]{FAD9D5} 0.000 \\
        SVM & 0.000 & 1.000 & \cellcolor[HTML]{FAD9D5} 0.490 & 0.000 & 1.000 & \cellcolor[HTML]{FAD9D5} 1.000 \\
        
        DNN & $<$0.001 & 1.000 & \cellcolor[HTML]{FAD9D5} 0.000 & $<$0.001 & 1.000 & \cellcolor[HTML]{FAD9D5} 1.000 \\

        \bottomrule
        \end{tabular}
    }
    \vspace{-3mm}
\end{table}

\subsection{Additional Variants of HsP}
\label{ssec:demo_variants}
\noindent
We study other types of HsP. To align these tests with the previous ones, we consider: different ssh-bruteforcing tools~(§\ref{sssec:demo_tools}), network-level manipulations~(§\ref{sssec:demo_bandwidth}), and system-level changes~(§\ref{sssec:demo_os}). Across these scenarios, we progressively relax the capabilities of the attacker. (The results are in Table~\ref{tab:demo_extra} in the Appendix~\ref{app:data})

\subsubsection{\textbf{Different tools}}
\label{sssec:demo_tools}
There are many ways to carry out ssh-bruteforcing attacks. Beyond \command{patator}{\small}, one can use well-known tools such as Medusa~\cite{medusa} or Hydra~\cite{hydra}. So, to reproduce a valid HsP, we consider the baseline models (trained on CICIDS17 and CICIDS18 on \command{patator {-}{-}persistent=1}{\small}) and test them against NetFlows generated by \textit{Medusa} and \textit{Hydra} (we used, once again, the network-traffic simulator in~\cite{verkerken2026concap} to do so). Note that these are all valid HsP: these tools do not require additional privileges, and they entail operations that, if successful, enable the attacker to achieve the same objective---violating an SSH server. The results indicate that, for \textit{Hydra}, two models (SVM and DNN) seem to be able to withstand such HsP on both datasets (\smamath{tpr}$>$0.890), but the remaining three models are always defeated (\smamath{tpr}$<$0.022); for \textit{Medusa}, the DNN is robust on CICIDS18 (\smamath{tpr}=0.879) and the RF is mediocre on CICIDS17 (\smamath{tpr}=0.658). Still, in all the remaining eight cases, the results indicate that the models are bypassed (the \smamath{tpr} is always below 0.334).

\subsubsection{\textbf{Changing the bandwidth}}
\label{sssec:demo_bandwidth}
We test an HsP similar to the adversarial strategy proposed in~\cite{catillo2024towards}. We envision an attacker with high privileges who attempts to bypass the ML-NIDS by inducing network-level changes through tampering with the Traffic Control utility. Such operations can be reproduced via the network simulator in~\cite{verkerken2026concap}, so we used it to generate NetFlows conforming to \command{patator {-}{-}persistent=1}{\small} (i.e., those included in CICIDS17 and CICIDS18) but by varying network-level parameters such as: \textit{latency}=(10ms, 25ms, 100ms), \textit{loss}=(0\%, 5\%), \textit{corrupt}=(0\%, 5\%), \textit{duplicate}=(0\%, 5\%). We assess all 24 combinations of these parameters, each time by launching \command{patator {-}{-}persistent=1}{\small}, capturing the corresponding traffic and labeling the NetFlows, and then testing them against the baseline models of CICIDS17 and CICIDS18. We find that only one model is defeated: the RF trained on CICIDS18 (\smamath{tpr}=0). In all other cases, the \smamath{tpr} ranges between 0.729 and 0.993. Hence, we conclude that such HsP have some impact (since the baseline \smamath{tpr} was always $>$0.999) but do not lead to a complete bypass---despite the fact that manipulating the network channel via Traffic Control requires root privileges (not needed for~§\ref{sssec:demo_tools}).

\subsubsection{\textbf{Different OS}}
\label{sssec:demo_os}
We test an ``almost-invalid'' HsP. We assume an attacker who, for some reason, can change the entire OS of the host they can control, switching from Ubuntu 16 (the one used in CICIDS17 and CICIDS18) to Ubuntu 18. This is an extreme HsP: the only way to do so in the real world, is if the attacker has physical access to the machine and replaces its OS. While we believe that real-world attackers would opt for other strategies before resorting to such a tactic, we still nonetheless tested the effects of such an HsP (which, we reiterate, assumes a ``stronger'' attacker than that in §\ref{ssec:demo_benchmark}). So, we used the network simulator in~\cite{verkerken2026concap}, thereby creating a machine with Ubuntu 18, and captured its traffic by launching \command{patator {-}{-}persistent=1}{\small}. Again, we found that these HsP are effective only against the RF trained on CICIDS18: in all other cases, the \smamath{tpr} remains above 0.75, indicating that these HsP are not very practical to bypass the ML-NIDS (which is only mildly impacted).

\begin{cooltextboxgreen}
\textsc{\textbf{Answer to RQ2:}} Some HsP can drop the \smamath{tpr} of an ML-NIDS from 1 to 0 by changing one character in the command launched by the attacker's controlled host. Even HsP leveraging a different tool but for the same malicious goal can lead to complete bypass. HsP requiring higher privileges are not necessarily more evasive. The effectiveness of HsP against specific ML models varies. 
\end{cooltextboxgreen}

In the Appendix~\ref{app:additional}, we expand our assessment by considering other attack types, models, and datasets.

\begin{figure*}[!t]
    \centering
     \begin{subfigure}[b]{0.6\linewidth}
         \centering
        \includegraphics[width=\linewidth]{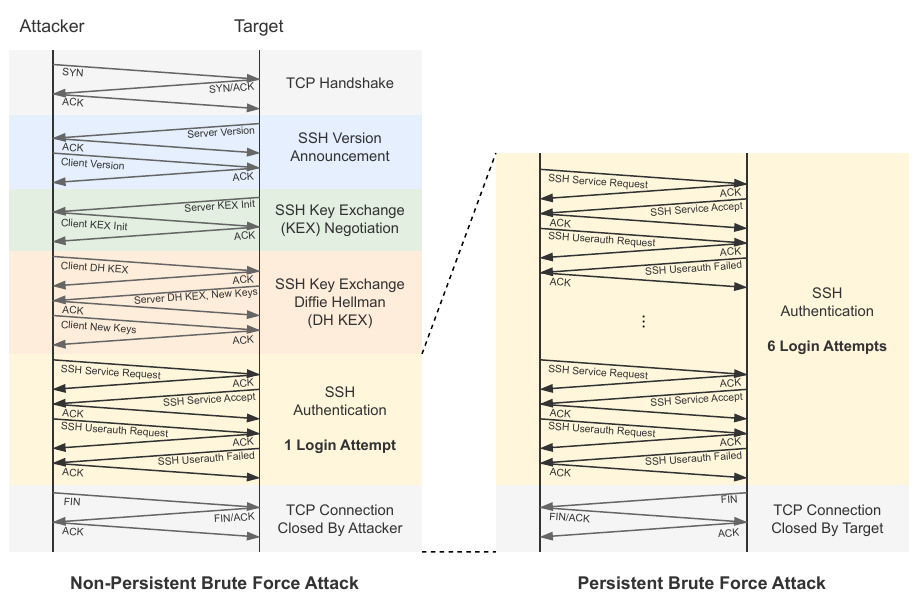}
        \vspace{-7mm}
        \caption{Packet-level. \textmd{By launching \command{patator -P=0}{\scriptsize} (left), only one login attempt is made. However, launching \command{patator -P=1}{\scriptsize} (right) leads to 6x more attempts. This difference is the ``packet-level perturbation'' induced by an HsP that changes \command{-P=0}{\scriptsize} to \command{patator -P=1}{\scriptsize} (i.e., a one-character change in the command launched by the attacker).}}
        \label{subfig:packets}
     \end{subfigure}
     \hfill
     \begin{subfigure}[b]{0.35\linewidth}
         \centering
         \includegraphics[width=\linewidth]{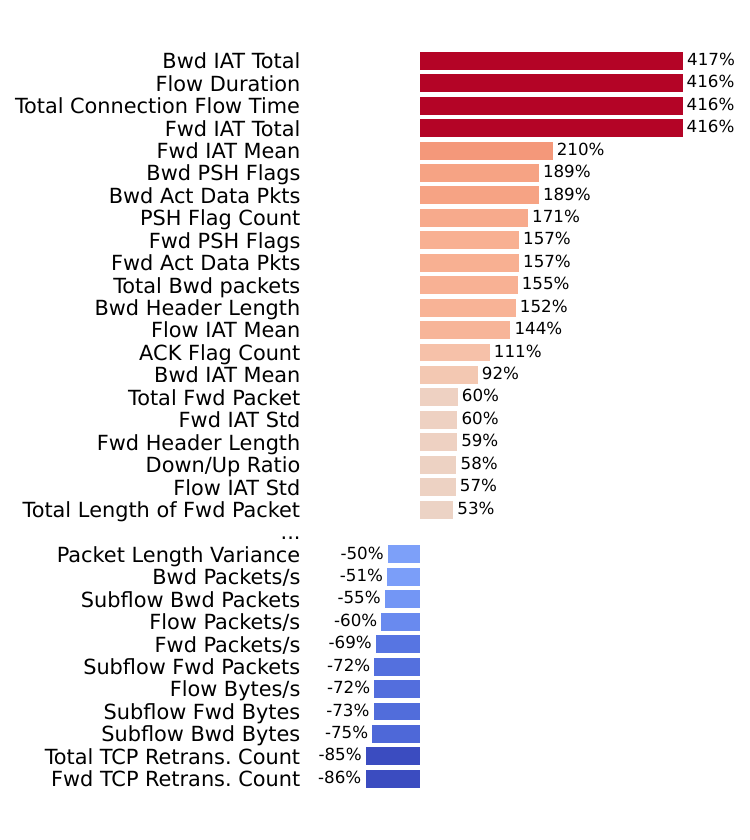}
         \vspace{-7mm}
         \caption{NetFlow-level. \textmd{In a the NetFlow-based feature space, an HsP switching  \command{-P=0}{\scriptsize} to \command{-P=1}{\scriptsize} leads to substantial differences (numbers in the plot are obtained by computing \command{-P=0}{\scriptsize}$~/~$\command{-P=1}{\scriptsize})}}
         \label{subfig:feature}
     \end{subfigure}
     \vspace{-3mm}
     \caption{\textbf{Low-level effects of a host-space perturbation.} \textmd{We show what happens when switching from \command{patator {-}{-}persistent=0}{\scriptsize} (-P=0) to \command{patator {-}{-}persistent=1}{\scriptsize} (-P=1). The former attempts a single login and closes the connection; the latter continues issuing login attempts until the SSH server closes the connection.}}
    \label{fig:casestudy}
    \vspace{-2mm}
\end{figure*}

\section{Case Study: Examination of SSH Patator}
\label{sec:casestudy}

\noindent
Our previous results raise a question: ``{\color{violet}Why is it that, for \command{patator}{\small}, changing a single character can lead to a complete bypass of the ML-NIDS?}'' We investigate this phenomenon by examining the data generated by the \command{patator}{\small} ssh-bruteforcing tool. The goal is twofold:  {\small \textit{(i)}}~provide more insights on how HsP ``mutate'' the resulting network traffic, producing different input samples that an ML-NIDS may struggle to classify---all with minimal effort from the attacker; and {\small \textit{(ii)}}~explain why HsP are tough to reproduce by directly manipulating pre-collected datapoints---especially in the feature space.

\subsection{Technical Analysis: what does \textmd{\command{patator}{\normalsize}} do?}
\label{ssec:theoretical}

\noindent
We explain how \command{patator}{\small} carries out its ssh-bruteforcing attempts.

At a high level, \command{patator}{\small} performs repeated SSH login attempts by trying combinations of usernames and passwords. It can be configured using several options, including the ``\command{persistent}{\small}'' flag. When enabled (\command{{-}{-}persistent=1}{\small}), \command{patator}{\small} continues attempting logins over a single TCP connection until the SSH server terminates it (by default, OpenSSH closes the connection after six failed attempts). In contrast, when disabled (\command{{-}{-}persistent=0}{\small}), \command{patator}{\small} initiates a new TCP connection for each login attempt. 
Fig.~\ref{subfig:packets} shows the operations in both modes, showing the different packet-level interactions between attacker and target with a default OpenSSH setup. While some communication patterns remain unchanged (e.g., the TCP handshake), setting \command{{-}{-}persistent=1}{\small} results in more~interactions. 

At first glance, one might assume that \command{patator {-}{-}persistent=1}{\small} simply leads to ``\smamath{\approx}6x more packets'' w.r.t. \command{patator {-}{-}persistent=0}{\small}. Yet, as we will show, the differences between enabling and disabling the ``persistent'' flag are more pronounced and less predictable.

\subsection{Real-world Quantitative Analysis}
\label{ssec:quantitative}

\noindent
To quantitatively compare \command{patator {-}{-}persistent=0}{\footnotesize} and \command{patator {-}{-}persistent=1}{\footnotesize}, we examine the NetFlows generated by each command. Doing so enables understanding why the ML models trained on one variant may not be able to classify the other ``HsP-driven'' variant.

We hence take the NetFlows generated in the real-world network (in §\ref{ssec:demo_real}), and we compare the various features (computed by CICFlowMeter, for which we used the fixed version in~\cite{engelen2021troubleshooting}) across the two considered variants of \command{patator}{\small}. We report in Fig.~\ref{subfig:feature} the ratio of the numerical features between \command{patator {-}{-}persistent=0}{\small} and \command{patator {-}{-}persistent=1}{\small}. Specifically, for each NetFlow feature, we aggregate all NetFlows of each variant of \command{patator}{\small} and then calculate the ratio. 

These results align with our explanation (§\ref{ssec:theoretical}). For instance, the \textit{FlowDuration} of \command{patator {-}{-}persistent=0}{\small} is much higher (over 4x) than that of \command{patator {-}{-}persistent=1}{\small}, because the latter issues multiple attempts (i.e., up to six) for each connection, whereas the former only issues one before closing the connection and initiating a new one (with a new attempt). At the same time, the longer duration also leads to a 75\% smaller \textit{Flow Bytes/s}. Put simply, these quantitative results show that these two variants of \command{patator}{\small} are substantially different at the feature level, explaining why an ML model trained over one variant may not be able to recognize the other variant as being of the same (malicious) class (i.e., an OOD-induced misclassification)). Still, depending on the decision boundaries learned by an ML model during its (unpredictable) training phase, the ML model may still be able to classify a variant of \command{patator}{\small} as malicious (e.g., the SVM trained only on \command{patator {-}{-}persistent=0}{\small} from the real-world setup was not fooled by \command{patator {-}{-}persistent=1}{\small}, see Table~\ref{tab:demo_real}).

To get a better understanding of the ``OOD'' potential of an HsP, we report in Fig.~\ref{fig:tsne} (in Appendix~\ref{app:data}) the t-SNE plot (inspired by~\cite{flood2024bad}) showing the distribution of the NetFlows produced by our two considered variants of \command{patator}{\small}, as well as that of CICIDS17 and benign NetFlows. We see that samples of \command{patator --persistent=1}{\small} (both ours and those in CICIDS17) form a single cluster (validating our implementation); but  \command{patator --persistent=0}{\small} is a clearly different cluster: therefore, these two variants belong to ``different distributions'', justifying why ML models may struggle identifying them.

Nonetheless, we make another observation: the differences at the NetFlow level ``exceed'' our expectations. For instance, \command{patator {-}{-}persistent=1}{\small} also has substantially higher values for the features related to the inter-arrival time (IAT) w.r.t. \command{patator {-}{-}persistent=0}{\small}; yet, there does not seem to be much correlation between these higher values (e.g., \textit{Fwd IAT Total} is 416\% higher, \textit{Bwd IAT Mean} is 92\% higher, \textit{Flow IAT Std} is 57\% higher). Moreover, the number of \textit{Ack Flag Count} of \command{patator {-}{-}persistent=1}{\small} is 111\% higher than that of \command{patator {-}{-}persistent=0}{\small}. Simply put, while we were aware that transitioning from one variant of \command{patator}{\small} to the other would have led to changes in multiple NetFlow features, we could not anticipate which features and to what extent they would be impacted. Therefore, \textit{we posit that replicating such differences via feature-space perturbations is (almost) impossible} (as we stated in §\ref{ssec:security}).

\subsection{Pitfall: Approximating HsP}
\label{ssec:pitfall}

\noindent
We elucidate what may happen when attempting to approximate HsP by directly manipulating pre-collected datapoints.

\subsubsection{\textbf{Manipulating the PCAP}}
\label{sssec:pcap}
Replicating the effects of our HsP by changing the packets within a PCAP trace is, we believe, not humanly feasible. Indeed, there are two cases here:
\begin{itemize}[leftmargin=*]
    \item \textit{Going from \command{patator {-}{-}persistent=0}{\small} to \command{patator {-}{-}persistent=1}{\small}:} this requires adding all the packets of the additional login attempts; but also deleting all the packets of the ``new connections''. 
    \item \textit{Going from \command{patator {-}{-}persistent=1}{\small} to \command{patator {-}{-}persistent=0}{\small}:} this requires to identify all packets of the ``persistent'' connections, delete them, and create new packets related to new connections. 
\end{itemize}
Practically realising both of the above requires applying precise changes to specific packets, including simulating new (or erasing existing) TCP handshakes and ACKs. All such changes must also account for the overarching network channel (e.g., there may be retransmissions). Hence, we do not see a humanly-feasible way to accurately reproduce a similar HsP by operating on a PCAP trace.

\subsubsection{\textbf{Manipulating the NetFlow features}}
\label{sssec:netflow}
Foreseeing the effects of an HsP on the NetFlow features is also challenging; evidence of this are the remarkably different values of the NetFlow features shown in Fig.~\ref{subfig:feature}. However, to provide evidence of ``what can go wrong'' if one attempts to (incorrectly) reproduce an HsP, we performed a simple experiment. According to our high-level analysis (§\ref{ssec:theoretical}),  launching \command{patator {-}{-}persistent=1}{\small} leads to \smamath{\approx}6x more packets (w.r.t. \command{patator {-}{-}persistent=0}{\small}) being exchanged. We ``naively'' simulated this by perturbing the feature space: we took the NetFlows generated by \command{patator {-}{-}persistent=0}{\small} (and \command{patator {-}{-}persistent=1}{\small}) in the real-world network, and multiply (divide) the number of packets by 6; we also did so for the number of bytes and updated all other dependent features, ensuring a correct feature vector that preserves domain constraints~\cite{sheatsley2021robustness}. Then, we submit these NetFlows, allegedly representing \command{patator {-}{-}persistent=1}{\small} (and \command{patator {-}{-}persistent=0}{\small}) to the ML models trained on the ``real'' variant of the corresponding \command{patator}{\small} command. Our rationale is that, if these ``fictitious'' NetFlows resemble that of the ``real'' \command{patator}{\small}, we should obtain the same results achieved in §\ref{ssec:demo_real}: a perfect \smamath{tpr} (see Table~\ref{tab:demo_real}). However, while this is true for two models (the DNN and the SVM), for three models (DT, HGB, RF) out of five, the \smamath{tpr} is \textit{always} 0. This indicates that using such NetFlows for security assessments of ML-NIDS in adversarial environments would be misleading, as they do not represent the attack that would occur in the real world.\footnote{\textbf{We do not cheat:} our intention is to use our prior knowledge of \command{patator}{\scriptsize} to derive such an estimate, which is based on what is discussed in §\ref{ssec:theoretical}. Clearly, by having knowledge of the quantitative differences between the two considered variants of \command{patator}{\scriptsize} (discussed in §\ref{ssec:quantitative}), such a ``naive'' approach would not make sense. Yet, using such knowledge to craft our feature-space perturbations would be cheating, but also pointless: such knowledge was obtained by generating \textit{real} network traffic, so it is pointless to reproduce the effects by changing the feature space, since it would be sufficient to use the appropriate PCAP trace and generate the corresponding NetFlows.} 

\begin{cooltextboxgreen}
\textsc{\textbf{Takeaway:}} Changing a single option (or parameter value) of a malicious command can lead to substantially different network activities---which achieve the same goal, but can cause misclassifications due to OOD (exploitable via HsP). Attempts to reproduce, or approximate, such changes by manipulating pre-collected datapoints is challenging and may result in misleading evaluations.
\end{cooltextboxgreen}
\section{Discussion and Clarifications}
\label{sec:discussion}

\noindent
We discuss our research, pointing out limitations~(§\ref{ssec:limitations}), making ethical remarks~(§\ref{ssec:ethics}), and clarifying intrinsic aspects of HsP~(§\ref{ssec:critical}).

\subsection{Limitations and Justifications}
\label{ssec:limitations}

\noindent
Our goal (§\ref{ssec:shortcomings}) is to substantiate the fact that prior work on ``adversarial ML attacks against ML-NIDS'' focused on perturbation techniques that are hard to realize by real-world attackers---who would rather opt for strategies captured by our notion of HsP.

To reach our goal, we have carried out an SLR (in §\ref{sec:slr}). While we have manually analysed \smamath{\approx}300 papers on this subject, and concluded that no prior work (aside from perhaps~\cite{catillo2024towards}) assessed the adversarial robustness of ML-NIDS to HsP, we acknowledge that our paper-selection methodology (based on reputable and recent works) may have led us to overlook some works that did employ techniques which fall in our definition of HsP. Still, even if this is true, it would not undermine our contribution: factually, most prior work on this subject applied perturbations in spaces that are not ``reachable'' by real-world attackers. We thus elucidate such a complementary ``adversarial-perturbation'' method to the ML-NIDS community.

Then, to provide evidence of the effects of HsP, we have carried out original experiments (in §\ref{sec:demo}). While such a demonstration entailed developing a total of 125 different models\footnote{Given by: 5 classifiers, trained over 5 different training sets (2 from the real world and 3 from benchmarks), repeated for 5 times for statistical robustness.} using both data from ``synthetic'' benchmarks and real-world network environments, we acknowledge that our primary evaluation (in §\ref{sec:demo} and §\ref{sec:casestudy}) mostly cover one specific class of HsP, namely, ssh-bruteforcing attacks. Even though we expanded our assessment (in the Appendix) with additional use cases (e.g., port scanning), covering all possible constillations of HsP (in terms of targeted ML-NIDS, considered network environment, and threat models is unfeasible and beyond the scope of a single paper. This, however, does not undermine our conclusions: HsP can impact ML-NIDS in ways that are not captured by perturbations applied to pre-collected datapoints.

Finally, and related to the point above, our evaluation encompasses only one category of ML-NIDS, i.e., those analysing NetFlow features. However, this also does not impact our conclusions: HsP do lead to network-level changes (at both {\small \textit{(i)}}~the packet level and {\small \textit{(ii)}}~the NetFlow level). Therefore, HsP are bound to affect any NIDS that falls into our definition (see §\ref{sec:background}). We do not claim that HsP are ``the only'', nor a ``guaranteed'', way to bypass ML-NIDS: therefore, our conclusions are not impacted in the slightest by the scope of our evaluation. Investigating the effects of HsP on all conceivable ML-NIDS would, therefore, just lead to a waste of compute.

\subsection{Ethical Considerations and Disclaimers}
\label{ssec:ethics}

\noindent
We do not seek to invalidate or diminish prior research. 

It is true that our findings/results reveal that the perturbations considered in prior work may not align with the real world. However, this does not undermine the contributions of prior work. First, because it is still possible to see the corresponding attacks as realistically plausible: this requires assuming a ``stronger'' attacker---who can, in some way, precisely control/manipulate the data acquired by devices such as routers or the NIDS itself. Second, because (some) prior works did warn that the considered perturbations may not be physically realizable (e.g.,~\cite{brockmann2025now}).
Third, because \textbf{it is thanks to such prior work that we were able to elaborate the notion of HsP}.
Therefore, prior work was fundamental for our contributions.

For the reason above, we do not seek to examine the experimental approaches adopted in prior work under a critical lens (e.g., to question the alignment between the experimental evaluation and the envisioned threat model). This would not be constructive: prior work (e.g.,~\cite{wang2022manda}) relied on methods that, to our knowledge, were the most appropriate ones to support their claimed contributions.

We obtained permission from the owners of the smart-home network to use their data for this research, and also to infect some hosts within their network with malicious tools and launch the corresponding attacks. However, for privacy, we cannot release the PCAP traces---but the NetFlows are included in our repository~\cite{repository}.

\subsection{Critical Remarks on HsP (and our response)}
\label{ssec:critical}

\noindent
In the spirit of a healthy scientific discourse, we report below, in a critique-and-response format, our stance on some legitimate critiques that a reader may have w.r.t. the general idea of HsP.

\textit{One can easily manipulate a PCAP trace (e.g., via scapy)} and send `perturbed' packets via tcpreplay. Therefore HsP are not the only way that attackers can use to evade ML-NIDS
\label{sssec:tcpreplay}
This is true (note: we never said manipulation of PCAP traces is not possible, nor that HsP are \textit{the only way} to evade ML-NIDS, see §\ref{ssec:limitations}). Attackers can manipulate packet structures directly, but with the risk of breaking protocols or leaving obvious artifacts (e.g., unnatural packet headers/payloads). Besides, even if the PCAP manipulation is functionally correct, the PCAP must still be ``replayed'' (by the host) and the resulting packets captured (by the router---i.e., the device which will forward the data to the ML-NIDS). Nonetheless, instructing a host to reproduce (via tcpreplay) an adversarially-manipulated PCAP trace is an operation that falls into our notion of HsP, as long as it is a permissible operation according to the given threat model (e.g., using tcpreplay requires root privileges, which the attacker may not have). \textit{Nevertheless, the data used for the assessment should be the one captured by the router---and not the one of the modified PCAP trace}. In other words, solely manipulating the packets included in a PCAP trace (assumed to be collected by the router/switch connected to the ML-NIDS) cannot exactly simulate the network activities, ``seen'' by the router, of the attacker-controlled host.

\textit{HsP are not adversarial perturbations!}
\label{sssec:notadversarial}
This is an opinion. It is true that the notion of ``adversarial perturbation'' typically denotes minimal changes in the feature space (e.g., changing one pixel~\cite{su2019one} or feature~\cite{alghuried2025evaluating}). However, over time, and especially in the security domain, such a notion has changed to encompass attacks wherein the adversary applies some changes to a given ``entity'' (e.g., an image, a STOP sign, a piece of malware, a website) which is then analysed by some ML model: such changes can be ``small' in the respective problem space, but ``large'' in the feature space (see, e.g.,~\cite{spacephish2022}). In the context of HsP, our notion simply denotes a way through which attackers attempt to reach their goal by performing ``different'' operations on their controlled hosts. The term ``different'' is what leads HsP to fall into the category of ``perturbations'': the attacker \textit{expects} that an ML-NIDS analyses the network communications of their controlled hosts, which is why the attacker attempts to reach their goal by perturbing/manipulating/modifying the commands they execute on their host. This is semantically not different from, e.g., applying perturbations that change malware source code (see~\cite{pierazzi2020intriguing}) or the HTML of a phishing webpage~\cite{spacephish2022}; there are even works whose ``adversarial perturbations'' lead to ``adversarial examples'' that are clearly discernible by humans~\cite{ji2025evaluating}. Therefore, while we acknowledge that HsP can lead to ``large'' changes to the feature space (see our OOD assessment in~§\ref{sssec:netflow}), HsP still qualify as ``adversarial perturbations'' (n.b.: the HsP used in our demonstration mostly entail changing a single character of a command!).

\textit{HsP require strong threat models to be staged!}
\label{sssec:strong}
This is not true. HsP do not require more knowledge/capabilities than those already envisioned in any given threat model (see §\ref{ssec:threat}): for instance, an attacker who wants to bypass an ML-NIDS via ``gradient-based'' perturbations does so because he/she expects the ML-NIDS to detect his/her malicious activities if he/she does not actively attempt to evade the ML-NIDS. We have also explicitly warned (in §\ref{ssec:security}) that one should be cautious when experimenting with HsP, as some commands may require privileges that do not conform to a given threat model (in which case, the HsP would not be valid for a security assessment). Finally, we note that all the HsP envisioned in this paper represent threat models wherein the targeted ML system is ``invisible'' to the attacker (according to~\cite{apruzzese2023real}).

\textit{HsP only apply to ssh bruteforcing!}
\label{sssec:only}
We primarily focused on ssh bruteforcing to provide a comprehensive demonstration (including both ML evaluation and subsequent data-level investigation) to facilitate understanding of HsP, and why they may work. However, HsP can be conceived for any type of malicious network activity. For instance, an attacker can carry out a port scan in a different way (as we described in §\ref{ssec:threat}), which would be an HsP {(which we test in the Appendix). An attacker can also instruct a botnet-infected host to perform its beaconing at different intervals, which would also be a valid HsP. An attacker can also use HsP to achieve the ``optimal'' feature-space perturbation that evades a given ML model. 

\section{Recommendations and Outlook}
\label{sec:recommendations}

\noindent
We derive lessons learned (§\ref{ssec:lessons}) and implications to the broader ML-NIDS domain (§\ref{ssec:implications}), and suggest avenues for future work~(§\ref{ssec:future}).

\subsection{Lessons Learned}
\label{ssec:lessons}

\noindent
We distill three lessons learned from our technical experiments. 

\textbf{Countermeasures.} If an ML-NIDS is bypassed via some HsP, such a vulnerability can be addressed via \textit{adversarial training}~\cite{shafahi2019adversarial}. Indeed, we have tested this experimentally: for all the ML models we developed in our evaluation (in §\ref{sec:demo}), we re-trained them by augmenting the training set with 80\% of the malicious NetFlows generated by any given HsP, and then testing the resulting model on the remaining 20\%: the models always obtained \smamath{tpr>0.999} (and we did not observe any decrease in the baseline performance). 

\textbf{Achieving robustness requires a system-wide mindset.} Our experiments show that: {\small \textit{(i)}}~minor changes to the command used to launch a network-related attack---such as going from \command{patator {-}{-}persistent=0}{\small} to \command{patator {-}{-}persistent=1}{\small}; and {\small \textit{(ii)}}~different implementations of the same network-related attack---such as using \textit{Medusa} or \textit{Hydra} instead of \textit{Patator}; lead to substantially lower detection rates by ML-NIDS trained on a single ``variant'' of the attack. Essentially, such small changes lead to OOD issues which cannot be countered via ML-centered solutions alone. Indeed, a single application of adversarial training cannot protect against all possible variants of HsP. This highlights that, to comprehensively assess whether a given ML-NIDS can withstand attacks of a specific class (e.g., ssh-bruteforcing attempts) it is necessary to test the ML-NIDS against multiple variants. Such variants must, in theory, encompass all the HsP that the envisioned attacker is allowed to launch---which requires system-wide considerations. Note, however, that some HsP can have no effect whatsoever on the network traffic (and, hence, on the ML-NIDS).\footnote{\textbf{(Negative result)} E.g., changing the \textit{threads} in patator (e.g., issuing \command{patator -T=1}{\scriptsize} or \command{patator -T=5}{\scriptsize}) does not lead to any network-level change if the attack targets a single ssh server. We tested this: the \scmath{tpr} never changed, and the NetFlows were not different.} We recommend to:
\ul{provide more specific threat models} (e.g., potentially stating, and justifying, what commands are, or not, allowed by the envisioned attacker); but also
\ul{study the various options that can be used to launch any given malicious command} to avoid carrying out potentially redundant experiments. Finally, to better optimize resource allocation, we recommend to \ul{examine the mapping between HsP and feature-level changes}: this way, one can train the ML-NIDS on a single command to provide robustness against multiple HsP variants of that command.

\textbf{Reproducing HsP.} From the viewpoint of a researcher, practically implementing HsP is challenging. This is because, to ensure correct implementation, it is necessary to operate directly on the host, and capture the corresponding network activities (otherwise, one may create inconsistencies, as shown in §\ref{ssec:pitfall}).
In our experiments, we adopted two approaches, depending on the use case.
\begin{itemize}[leftmargin=*]
    \item \textit{Complete ``real-world'' simulation.} If the researcher has complete control over the network environment from which the network-traffic data is generated, then exploring different HsP is straightforward. In our case (§\ref{ssec:demo_real}), we simply launched a different command on the hosts, captured the traffic, labeled it appropriately, and used these datapoints in our ML-focused experiments. However, obtaining access to such a setup requires some effort. A viable alternative is using virtual machines to create a virtual network (some existing datasets have been created in this way~\cite{flood2024bad}); or use specific network-simulation tools (e.g.,~\cite{hariri2024rl,verkerken2026concap}).

    \item \textit{Reliance on ``benchmark'' datasets.} If one wants to assess the robustness of ML-NIDS developed by using data included in a benchmark dataset (e.g., CICIDS17), then it is necessary to adopt a different approach. The data in a benchmark dataset is captured in a specific network; hence, when creating the HsP, the host on which the commands are run must resemble the host used to create the data included in the benchmark. In our case (§\ref{ssec:demo_benchmark}), we studied the documentation of CICIDS17/18 and thoroughly examined the data contained in these datasets to derive an appropriate ``configuration'' that we could use (alongside the network-simulation tool in~\cite{verkerken2026concap}) to create realistic HsP.    
\end{itemize}
Put differently, implementing HsP requires \textit{expertise in networking}.

\subsection{Implications of our Research}
\label{ssec:implications}
\noindent
We identify three major implications of our findings.

First, our SLR revealed that prior works examining ``adversarial ML attacks'' against ML-based NIDS carried out their evaluations by considering perturbations that are -- while not necessarily unrealistic -- hard to practically realise. From the viewpoint of an attacker, the main obstacle to overcome in order to apply such ``traditional'' perturbations is \textit{finding a way to manipulate datapoints that are collected on dedicated, and highly-secure, appliances} (e.g., the router, or the NIDS itself). Note: we are not claiming that practically evading ML-NIDS cannot be done via, e.g., gradient-based strategies whose results are applied in the feature space; rather, we argue that doing so requires assuming that the attacker has also compromised hosts beyond those that they can remotely control. Hence, future work seeking to study such forms of adversarial ML attacks should justify how the envisioned attacker can perturb pre-collected datapoints.

Second, our experimental evaluation indicated that minimal changes to the malicious commands launched by the attacker can induce substantial alterations in the features analysed by an ML-NIDS, which can lead to evasion. Given the lack of work examining these classes of ``adversarial ML strategies'', we hence endorse future work to also explore this dimension when assessing the robustness of ML-NIDS. At the same time, we endorse future work to be explicit in providing the commands/options/source-code used to launch any attack included in an experimental evaluation.

Third, the concept of HsP opens the debate of ``when do adversarial perturbations stop being an ML-security problem and become a system-security problem?'' Indeed, as we discussed (in §\ref{sssec:notadversarial}), the concept of adversarial perturbations has evolved over the years. Given that ML models are now increasingly integrated into operational systems (including NIDS~\cite{apruzzese2023sok}), we believe that these two domains (ML and system security) should not be disjoint. We posit that studying the security of ML-NIDS should start from a fundamental question: ``\ul{in what ways can the attacker, within their problem space, \textit{feasibly} influence the data analysed by the ML model?}'' Indeed, strengthening the considered ML model only against perturbations applied to pre-collected datapoints may lead to a false sense of security if the attacker can achieve evasion by, e.g., merely change a character of a malicious command.

\subsection{Future Work}
\label{ssec:future}

\noindent
There are several aspects that can be addressed in future work. 

First, finding ways to map packet-level (or feature-level) changes to HsP (which is orthogonal to our recommendation in §\ref{ssec:lessons}). For instance, how can an attacker, via HsP, realise the gradient-based attack that would induce a misclassification despite a minimal change in the feature representation of a given sample? This would lead to identifying HsP that are guaranteed to evade the ML-NIDS.

Second, future research can investigate how HsP affect different classes of ML-NIDS, such as those analysing only PCAP traces~\cite{wang2022manda} (which may not be feasible if the traffic is encrypted, and NetFlows would be more informative~\cite{vormayr2020my}), or provenance logs~\cite{cheng2024kairos}, or even network graphs~\cite{venturi2023arganids,song2024madeline}. Certain HsP may have no impact whatsoever on the datatype analyzed by some ML-NIDS. For instance, we can reasonably infer that graph-based NIDS may not be impacted by the HsP focused on \command{patator}{\small}, because the hosts involved in the exchange would still be the same, meaning that the network graph should not be altered. Yet, if the attacker expects the NIDS to rely on graph-based algorithms, the attacker may rely on HsP that alter the network graph (e.g., by triggering connections with other hosts in the network). We believe that our notion of HsP should inspire future work to investigate which types of ML-NIDS can be used to cover potential ``blind spots'' of certain classes of detectors.

Third, HsP are a way to achieve evasion at test/inference time. Our follow-up adversarial training experiments (§\ref{ssec:lessons}) showed that, by retraining an ML model on the data generated via HsP, it is possible to make such an ML model robust against the specific HsP. However, exhausting all possible combinations of HsP may lead to ML models that, while being effective against the broad class of attacks related to the HsP, may not be robust against other types of attacks. In a sense, this could be a form of ``self poisoning'', wherein one trades specificity for generalizability. Future work can therefore explore whether there are any tradeoffs: for instance, only a subset of all possible HsP is sufficient to cover the entire spectrum of HsPs of a given attack, and generating new training data would lead to detrimental effects from a generalizability standpoint.

\section{Conclusions}
\label{sec:conclusions}

\noindent
We elucidated a blind spot in prior literature on ``adversarial ML attacks'' against ML-based NIDS.

The potential impact of our proposed host-space adversarial perturbations (HsP) is a call to action. Real attackers favor cheap and simple strategies. 
Future research should shed more light on this, yet another, security risk of ML-NIDS.

\begin{acks}
    \noindent
    We thank the anonymous reviewers for their feedback.
This research was partially supported by Hilti, and by the BOSA-AIDE project funded by the Belgian SPF BOSA with reference number 06.40.32.33.00.10. This work originated from a research stay at the University of Liechtenstein and was supported by the Research Foundation – Flanders (FWO) under grant number V450224N.
\end{acks}

\bibliographystyle{ACM-Reference-Format}



\appendices
\section{On the ``spaces'' of perturbations}
\label{app:perturbations}

\noindent
To further illustrate the differences between the ``spaces'' of adversarial perturbations, we describe three exemplary scenarios that, despite involving domains orthogonal to NIDS, enable a better understanding of the crux tackled by our work.

\subsubsection*{\textbf{Attacks vs Autonomous Driving Cars}}
\label{ssapp:driving}
Assume a car that relies on ML to analyse the surrounding environment so that proper driving decisions can be made. An exemplary use case is an ML model that must recognize \textit{traffic signs}. 
In this case, the ``feature space'' is represented by the pixels of the images acquired by the sensors of the car (e.g., cameras), which are provided as input to the ML model to discern whether {\small \textit{(i)}}~the image contains a traffic sign, and {\small \textit{(ii)}}~if so, which sign it is---so that the car can react accordingly.

To introduce a perturbation in such a feature space (i.e., direct pixel manipulation of the digitally-acquired image), the attacker can \textbf{(A)} tamper with the software embedded in the car's system---which is doable, but impractical unless the attacker has compromised the systems of the car's manufacturer, or manipulated the systems of the targeted car~\cite{woitschek2021physical}. 
The attacker, however, can also \textbf{(B)} manipulate the traffic sign \textit{in the physical world}. To do this, the attacker can, e.g., tamper with a specific traffic sign, so that whenever some sensors capture images of such a traffic sign, the resulting image will (ideally) fool the classification of the ML model~\cite{eykholt2018robust,song2023discovering}.\footnote{This example is useful to explain a ``host-space'' perturbation: instead of manipulating the image, which is acquired and processed by devices that are not under the attacker's control (i.e., a perturbation to a piece of data), the attacker creates the ``perturbation'' by interacting with the physical world (i.e., which, in our case, is the attacker's host).}

N.b.: \textbf{(A)} and \textbf{(B)} denote two different threat models where the attacker's goal is the same, but the capabilities intrinsically change due to a semantically different problem space: in \textbf{(A)}, the problem space overlaps with the feature space; in \textbf{(B)}, there is no overlap.

\subsubsection*{\textbf{Attacks vs Malware Detectors}}
\label{ssapp:malware}
Assume a malware detector that relies on ML to analyse some applications and determine whether they are malicious or not. An exemplary use case is the DREBIN detector for Android malware~\cite{arp2014drebin}: as input, DREBIN receives a feature vector of thousands of binary features, each related to a specific functionality of the corresponding Android app (e.g., permissions). 

To \textit{reliably} introduce perturbations in such a feature space, the attacker must have access to the preprocessing mechanism that extracts the feature vector (which will be analysed by the ML model, i.e., DREBIN), or otherwise be able to change its output. Doing so means that the attacker has root access of either {\small \textit{(a)}}~the Android device executing the anti-malware app, or {\small \textit{(b)}}~of a remote security system. Both of these cases are possible, but assume a very powerful attacker who has already compromised the security system.

Alternatively, the attacker can \textit{manipulate the source-code} of the (malicious) Android app, so that the resulting feature vector will have values that bypass the ML model~\cite{pierazzi2020intriguing,lucas2024training}. This is easier to do by an attacker, since they have complete control over the source code of the piece of malware that they develop; however, in doing so, the attacker must be careful not to implement changes that would hinder the malicious functionality of the resulting app.

\subsubsection*{\textbf{Attacks vs Phishing Website Detectors}}
\label{ssapp:phishing}

To counter phishing websites mimicking benign websites, state-of-the-art solutions rely on ML models for visual recognition. For instance, PhishIntention~\cite{liu2022inferring} extracts the logos from the screenshot of a given webpage, and then uses ML to determine if such logos resemble those of a well-known brand (e.g., PayPal); then, if a match is found, the systems checks if the domain of the analysed webpage matches the domain of the well-known brand (ideally, phishing webpages resemble well-known brands, but are hosted under different domains). In this case, the feature space is represented by the image of a logo (extracted from the screenshot of a webpage).

To introduce perturbations in such a feature space, the attacker simply needs to visually alter the logo in the phishing webpage---on which they have, by definition, complete control. For instance, the attacker can take the logo image, change some pixels, and stitch the new ``adversarial'' logo onto the phishing webpage~\cite{hao2024doesn}. 

This is an exemplary case in which \textit{the attacker may have complete control on the feature space}, because the attacker's problem space ``overlaps'' with the ML-model's feature space. However, there are other types of phishing website detectors (reliant, e.g., on the analysis of information extracted from the HTML) that do not allow an attacker to tamper with the feature space---unless the attacker has access to the internal workflow of the targeted phishing detection system. We point the interested reader to~\cite{kotzias2023scamdog,spacephish2022}.

\begin{cooltextboxgreen}
\textsc{\textbf{Takeaway.}} Adversarial perturbations can be applied in many ways and in many ``spaces''. Yet, depending on the specific application domain, introducing some perturbations may be challenging for real-world attackers. Hence, \textit{it is crucial to identify the ``problem space'' in which the attacker is allowed to operate}. In such a way, it is possible to define what actions an attacker can take to reach their underlying goal---and, hence, simulate plausible adversarial strategies, studying their effects, and devise countermeasures.
\end{cooltextboxgreen}

\begin{table}[t]
    
    \caption{\textbf{Additional security assessments.} \textmd{\footnotesize We report the \scmath{tpr} achieved by the baseline models trained on benchmarks (CIC17, CIC18) against the NetFlows conforming to different host-space perturbations}}
    \label{tab:demo_extra}
    \vspace{-3mm}
    \centering
    \resizebox{0.95\columnwidth}{!}{
        \begin{tabular}{
        l?cc|cc|cc|cc
        }
        \toprule
        \multirow{2}[1]{*}{Model} & \multicolumn{2}{c}{Hydra} & \multicolumn{2}{c}{Medusa} & \multicolumn{2}{c}{Network change} & \multicolumn{2}{c}{OS change} \\

        & CIC17 & CIC18 & CIC17 & CIC18 & CIC17 & CIC18 & CIC17 & CIC18 \\
        \midrule
        DT & 0.019 & 0.000 & 0.000 & 0.000 & 0.743 & 0.729 & 0.750 & 0.750 \\
        RF & 0.018 & 0.000 & 0.658 & 0.000 & 0.993 & 0.000 & 0.849 & 0.000 \\

        XGB & 0.021 & 0.000 & 0.005 & 0.000 & 0.883 & 0.729 & 0.956 & 0.750 \\

        SVM & 0.890 & 0.895 & 0.214 & 0.334 & 0.749 & 0.733 & 0.750 & 0.750 \\
        DNN & 0.941 & 0.902 & 0.221 & 0.879 & 0.822 & 0.749 & 0.768 & 0.750 \\
        \bottomrule
        \end{tabular}
    }
    \vspace{-3mm}
\end{table}

\section{Data Collection and Preprocessing}
\label{app:data}
\noindent
We explain the data collection and preprocessing for our tests. We also report some experimental results (Table~\ref{tab:demo_extra} and Fig.~\ref{fig:tsne}).

\begin{figure}[t]
    \centering
    \includegraphics[width=0.85\columnwidth]{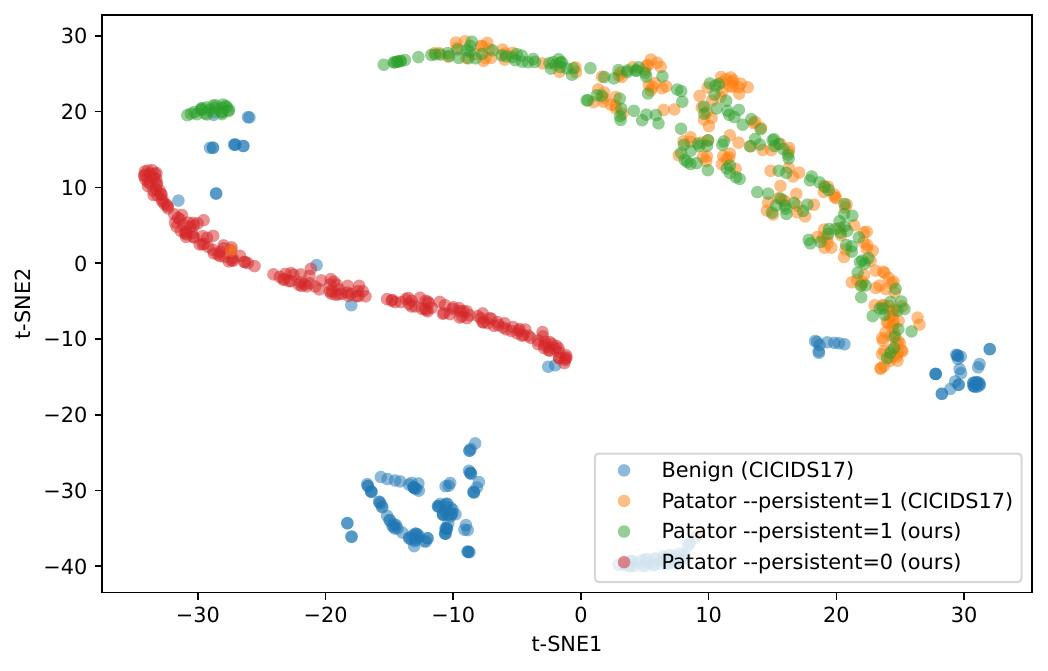}
    \vspace{-4mm}
    \caption{OOD verification. \textmd{We use a tSNE plot to visualize the distribution of NetFlows generated by \command{patator}{\scriptsize} (as well as benign ones).}}
    \label{fig:tsne}
\end{figure}

\subsection{Real-world Smart-home network capture}
\label{sapp:smarthome}

\noindent
The network environment contains 40--50 physical devices. These include smartphones, laptops/desktops, gaming consoles, and various IoT devices (e.g., smart speakers and lightbulbs). All devices connect to a router via a WiFi 5 or 2.4 interface. The router has an Internet connection with 50 Mbps download and 5 Mbps upload. The router is a FritzBox 7590, and a (local) Raspberry Pi4 with PiHole serves as DNS server, forwarding queries to Cloudflare. Devices on the network receive security patches and their owners are security experts, so it is reasonable to treat the traffic as ``benign'' (and even if some traffic is ``malicious'', it belongs to a different class than \command{patator}{\small}, so our conclusions remain unaffected). To implement our attack we used two laptops, both running Ubuntu 18.04: the host from which the attack was launched mounts an Intel i7-7700H with 32 GB of RAM; whereas the host acting as the target mounts an Intel N4100 with 8 GB of RAM. 

For our experiments, three PCAP traces were captured: two consisting of benign background traffic, and one related to the \command{patator}{\small} SSH brute-force attacks. The background traffic was first captured on 6 November 2023 and a second time on 26 August 2024---the same day the malicious attacks were executed and captured. The benign traffic from the 26 August capture totals 6 GB for 8M packets (yielding nearly 50K NetFlows), whereas the attack capture measures 45MB for 241k packets (corresponding to 8k NetFlows). The (benign) trace captured in November 2023 is 17GB for 17M packets, resulting in 30k NetFlows (extracted via CICFlowMeter)

\subsection{Handling the CICIDS17/18 Datasets}
\label{sapp:benchmark}

\noindent
The CICIDS17 and CICIDS18~\cite{sharafaldin2018toward} benchmark datasets have been known to have various issues for years (see~\cite{flood2024bad, engelen2021troubleshooting, liu2022error}). We are aware of this, which is why we always used the ``fixed'' version of these datasets, following the methods discussed in~\cite{engelen2021troubleshooting} (for CICIDS17) and~\cite{liu2022error} (for CICIDS18) to ensure that our experimental setup was correct and that labelling issues, if present, were minimal. After downloading the source PCAP data for these datasets, we extracted the NetFlows via the (fixed) version of CICIFLowMeter, and applied the labeling logic in~\cite{liu2022error,engelen2021troubleshooting}. We removed the ``attempted'' NetFlows, since they are not useful for our purposes.

Then, we cleaned the resulting NetFlows from features that have historically been known to create shortcuts during the training stage of ML models. Specifically, following the recommendations of prior work~\cite{apruzzese2022sok, arp2022and, d2022establishing}, we remove the following features: ``id'', ``Flow ID'', ``Src IP'', ``Dst IP'', ``Timestamp'', ``FWD Init Win Bytes'', and ``Bwd Init Win Bytes''. Then, we associate the source and destination ports to the IANA port-type categories (and then encoded as 0, 1, or 2 for respectively \textit{well-known}, \textit{registered}, and \textit{dynamic}). Finally, we sanitized all missing values and removed any duplicate NetFlow.

\section{Additional Experiments}
\label{app:additional}
\noindent
We describe additional experiments we have carried out to further expand our assessment and provide a broader understanding of the potential effects of HsP. Particularly, we provide further evidence that HsP can lead to ``OOD'' samples that cause misclassifications.

{\setstretch{0.7}
\textbox{{\footnotesize \textbf{Disclaimer.} The spectrum of possile attacks and corresponding HsP is virtually infinite. We cannot cover all such possibilities. Here, we merely want to provide additional insight on the potential of HsP-driven attacks to bypass ML-NIDS.}}}

\subsection{HsP against DeepLearning-based NIDS}
\label{sapp:deep}
\noindent
In the main paper, we considered ``traditional'' detectors created using the scikit-learn toolkit. Here, we expand our assessment by considering additional detectors based on more complex (and recent) deep-learning (DL) techniques and frameworks.

\textbf{Objective and Method.} 
We want to see if ``advanced'' DL-based detectors can also be fooled by the same HsP considered in our primary experiment (§\ref{ssec:demo_benchmark}).
We consider three classes of DL-driven detectors. Two are inspired by~\cite{wei2023xnids}, which carried out evaluations considering a DL-based \textit{autoencoder} (drawn from~\cite{mirsky2018kitsune,shone2018deep}), as well as a \textit{recurrent neural network} (drawn from~\cite{jan2020throwing,yin2017deep}). The third is an application of a \textit{large-language model} (LLM), inspired by~\cite{luay2025multimodal, verkerken2026concap}.

\textbf{Implementation.} 
For the autoencoder (AE), we use PyTorch to replicate the 6-layer AE in~\cite{wei2023xnids} (also used in~\cite{mirsky2018kitsune}). For the recurrent neural network (RNN), we tried to replicate long-short-term-memory method in~\cite{jan2020throwing}, but we ultimately opted for a gated-recurrent unit classifier because it was faster to train. 
For the LLM, we use ChatGPT 5.4 (the best available as of March 2026). We use the same one-shot prompting technique in~\cite{verkerken2026concap} where the LLM is told to classify NetFlows in malicious/benign while being provided with ``training'' data to so that it can determine some boundaries between benign and malicious datapoints; we also tried a full zero-shot prompting strategy, which did not yield good results (the LLM classifies everything as benign). The code for both the RNN and the AE, as well as the prompts for the LLM, are in our repository~\cite{repository}.

\textbf{Evaluation.}
To align this experiment with that in our main paper (in §\ref{ssec:demo_benchmark}), we consider the (fixed) CICDS17 dataset, and consider its benign samples alongside the malicious ones of \command{patator {-}{-}persistent=1}{\small} to develop our DL-based classifiers; we split this dataset in train:test with an 80:20 split. Then, we submit the HsP-variant of the considered attack (\command{patator {-}{-}persistent=0}{\small}) and see how the DL-based classifiers react. We report the available results in Table~\ref{tab:extra_deeplearn}. We can see that these more complex models have a very good baseline performance (high \smamath{tpr} with low \smamath{fpr}) but cannot recognize our HsP as malicious. Note that we do not seek to generalize this conclusion: we simply show the impact of HsP on other types of classifiers.

\begin{table}[t]
    
    \caption{\textbf{Experiments on additional DL-based classifiers.}} 
    \label{tab:extra_deeplearn}
    \vspace{-3mm}
    \centering
    \resizebox{0.45\columnwidth}{!}{
        \begin{tabular}{
        l?ccc
        }
        \toprule
        & \smamath{tpr} (baseline) & \smamath{tpr} (HsP) & \smamath{fpr} \\
        \midrule
        AE~\cite{wei2023xnids} & 0.9983 & 0.0000 & 0.0026 \\
        RNN~\cite{wei2023xnids} & 0.9899 & 0.0000 & <0.0001 \\
        LLM~\cite{verkerken2026concap} & 0.9949 & 0.0000 & <0.0001 \\
        \bottomrule
        \end{tabular}
    }
    \vspace{-4mm}
\end{table}

\subsection{HsP entailing different Attacks}
\label{sapp:attacks}
\noindent
Our main paper (in §\ref{sec:demo}) mostly considered HsP focusing on ssh-bruteforcing attacks (e.g., \command{Medusa}{\small}, \command{Hydra}{\small}, or \command{patator}{\small}). Here, we expand our assessment by showing other use-cases of HsP.

\textbf{Attacks.} We consider 
\textit{fuzzying} (via \command{wfuzz}{\small}~\cite{wfuzz}), \textit{port-scanning} (via \command{nmap}{\small}), and \textit{ddossing} (via \command{slowloris}{\small}~\cite{slowloris}). We consider these attacks because they align with the setup of our primary experiments (in §\ref{sec:demo}): the (fixed) CICIDS17 dataset contains port-scanning (via \command{nmap}{\small}), ddossing (via \command{slowloris}{\small}), and fuzzying (implemented via custom-made scripts for web attacks---which can be replicated via \command{wfuzz}{\small}). So, our choice of attacks allows creating valid HsP that can be assessed against the classifiers considered in our main paper (but trained on the ``baseline'' variant of the attack included in CICDS17). We assume an attacker with the same knowledge and capabilities as that in our primary evaluation (§\ref{ssec:demo_benchmark}), but the goal is clearly different (i.e., fuzzying/ddossing/portscanning instead of ssh-bruteforcing).

\textbf{Implementation.} We consider the five classifiers (DT, SVM, RF, DNN, HGB) and train them on 80\% of the benign NetFlows of CICIDS17, and 80\% of the NetFlows available for the fuzzying, ddos, and portscanning attacks in CICIDS17; we then test these classifiers on the remaining 20\% to assess their performance in the absence of HsP. Then, we craft HsP by using the same network-simulator tool used in our main paper~\cite{verkerken2026concap}, configured in the same way used for the experiments on CICIDS17 (§\ref{ssec:demo_benchmark}). Specifically, for portscan, we consider several variants of \command{nmap}{\small}; for fuzzying, we consider five variants of \command{wfuzz}{\small}; for ddos, we consider six variants of \command{slowloris}{\small}. We report the specific commands in the Supplementary material. 

\textbf{Evaluation.}
The results are shown in Table~\ref{tab:extra_atk}. For ddossing, the classifiers are always defeated. For fuzzying, DT, RF, and HGB retain limited recall, while SVM and DNN collapse completely. For portscanning, there are cases in which the HsP have little effect, because they were similar to the variant of \command{nmap}{\small} used in CICIDS17 (which, e.g., considered the options \command{{-}sS}{\small} or \command{{-}sT}{\small}~\cite{sharafaldin2018toward}, and was likely launched with the default \command{-T3}{\small}). It is hence expected that such HsP do not have a substantial impact on the classifiers. However, some variants still successfully bypass our detectors.

\subsection{Assessments on a different Benchmark}
\label{sapp:dataset}
\noindent
In our paper, we tested HsP against ML models trained on data from a real-world network (§\ref{ssec:demo_real}) and on two well-known benchmark datasets: CICIDS17 and CICIDS18 (§\ref{ssec:demo_benchmark}). Here, we show the applicability (and impact) of HsP on a completely different setup.

\textbf{Setting and Threat Model.} We consider a network resembling a typical Internet-of-Things environment. It encompasses various smart devices, such as smart devices (e.g., smart bulbs or smart switches, or smart doorbells), cameras, but also laptops and servers. The network is protected by an ML-NIDS that receives network-traffic data collected from the router in NetFlow format; the ML-NIDS is trained to detect some DDoS and SSH-bruteforcing attacks. We assume an attacker who has infiltrated in this network by compromising one host, and seeks to launch DDoS attacks from the inside, or attempt SSH bruteforcing attempts (the attacker's knowledge and capabilities hence align with the assumptions in our main paper §\ref{sec:demo}). The attacker seeks to reach their goal by means of HsP, hoping to bypass the detection of the ML-NIDS.

\textbf{Experimental setup.} To implement such a use case, we consider the CIC-BCCC-NRC TabularIoTAttack-2024 benchmark dataset~\cite{sasi2024efficient}; particularly, we focus on the partition discussed in ACI-IoT-2023~\cite{nack2024aci}. This partition contains labeled NetFlow data (generated via CIC-FlowMeter 3.0) of an IoT network resembling that of our envisioned threat model. Among the malicious NetFlows, some are generated via \command{hydra}{\small} (which is an ssh-bruteforcing tool~\cite{hydra}---also used in our main paper~§\ref{sssec:demo_tools}) and some are generated by \command{slowloris}{\small} (which is a tool for launching DDoS attacks~\cite{slowloris}). So, after {\small \textit{(i)}}~studying the documentation in~\cite{nack2024aci}
we {\small \textit{(ii)}}~use the network traffic generator in~\cite{verkerken2026concap} to replicate a setup aligning with that in~\cite{nack2024aci}---such that the hosts involved in the malicious communications are aligned with that of the authors of~\cite{nack2024aci}.\footnote{E.g., we set a network channel of 100Mbit/s with no packet loss because this is typical in a relatively-small network environment, especially given that our attacks entail internal-to-internal communications.} Then, we {\small \textit{(iii)}}~use our setup to launch HsP related to our two considered attacks. Specifically:
\begin{itemize}[leftmargin=*]
    \item  for \command{hydra}{\small}, we used: \command{hydra .L /\$Path/top-usernames-shortlist.txt -P /\$Path/2023-200\_most\_used\_passwords.txt ssh://\$IP -t 10}{\small}
    
    \item  for \command{slowloris}{\small}, we change the default script (available in~\cite{slowloris}) to use HTTPS mode, i.e., by launching \command{python3 ./slowloris.py {-}{-}https}{\small}
    
\end{itemize}
N.b.: the objective of this proof-of-concept experiment is to investigate the impact of some HsP on a different network environment.\footnote{Unfortunately, due to lack of documentation, we were unable to identify the exact command used to launch \command{hydra}{\scriptsize} and \command{slowloris}{\scriptsize} in~\cite{nack2024aci}. Nonetheless, we have reason to believe that our ``custom'' commands are different from those used by the creators of~\cite{nack2024aci}. For instance, by default, \command{hydra}{\scriptsize} sets {\command{-t 16}{\scriptsize} whereas we use \command{-t 10}{\scriptsize} (see~\cite{hydra})}}

\textbf{Evaluation.} We train the same classifiers used in our main paper on the benign and malicious samples of our considered attacks in~\cite{nack2024aci}, reserving 20\% of the available NetFlows for testing purposes; then, we test them on our HsP variants of these attacks. The results are in Table~\ref{tab:extra_benchmark}. Our classifiers attain near-perfect \smamath{tpr} against the ``baseline'' variant of each of our attacks, while also exhibiting low \smamath{fpr} for the tree-based models and somewhat higher \smamath{fpr} for SVM/DNN. However, against our HsP-variants of \command{hydra}{\small}, only DT is not affected: all other classifiers recognize these NetFlows as benign. For \command{slowloris}{\small}, all classifiers are defeated. We can therefore conclude that HsP can be dangerous even in a different setup.

\begin{table}[t]
    
    \caption{\textbf{Experiments on more attacks.} \textmd{\footnotesize We test HsP related to portscanning (\command{nmap}{\scriptsize}), fuzzying (\command{wfuzz}{\scriptsize}), and DDoS (\command{slowloris}{\scriptsize}). These attacks are included in CICIDS17.}}
    \label{tab:extra_atk}
    \vspace{-3mm}
    \centering
    \resizebox{0.95\columnwidth}{!}{
        \begin{tabular}{
        l?ccc|ccc|ccc
        }
        \toprule
        \multirow{2}[1]{*}{Model} & \multicolumn{3}{c|}{Wfuzz} & \multicolumn{3}{c|}{Nmap} & \multicolumn{3}{c}{Slowloris} \\
        & \smamath{tpr} (baseline) & \smamath{tpr} (HsP) & \smamath{fpr} & \smamath{tpr} (baseline) & \smamath{tpr} (HsP) & \smamath{fpr} & \smamath{tpr} (baseline) & \smamath{tpr} (HsP) & \smamath{fpr} \\
        \midrule
        DT & 0.9524 & 0.2000 & 0.0476 & 0.9985 & 0.6912 & 0.0022 & 0.9993 & 0.0000 & 0.0017 \\
        RF & 1.0000 & 0.2000 & 0.0476 & 0.9996 & 0.6324 & 0.0004 & 1.0000 & 0.0000 & 0.0000 \\
        HGB & 0.9524 & 0.2000 & 0.0476 & 0.9996 & 0.7059 & 0.0007 & 1.0000 & 0.0000 & 0.0000 \\
        SVM & 1.0000 & 0.0000 & 0.0000 & 0.9922 & 0.5735 & 0.0101 & 0.9990 & 0.0000 & 0.0003 \\
        DNN & 1.0000 & 0.0000 & 0.0000 & 0.9970 & 0.6324 & 0.0037 & 0.9997 & 0.0000 & 0.0007 \\
        \bottomrule
        \end{tabular}
    }
    \vspace{-4mm}
\end{table}

\begin{table}[!h]
    \vspace{-2mm}
    \caption{\textbf{Experiments on different benchmark.} \textmd{\footnotesize We test HsP of \command{hydra}{\scriptsize}~\cite{hydra} and \command{slowloris}{\scriptsize}~\cite{slowloris} on the~\cite{nack2024aci} section of CIC-BCCC-NRC TabularIoTAttack-2024~\cite{sasi2024efficient}.}}
    \label{tab:extra_benchmark}
    \vspace{-3mm}
    \centering
    \resizebox{0.85\columnwidth}{!}{
        \begin{tabular}{
        l?ccc|ccc
        }
        \toprule
        \multirow{2}[1]{*}{Model} & \multicolumn{3}{c|}{Hydra~\cite{hydra}} & \multicolumn{3}{c}{Slowloris~\cite{slowloris}}  \\
        & \smamath{tpr} (baseline) & \smamath{tpr} (HsP) & \smamath{fpr} & \smamath{tpr} (baseline) & \smamath{tpr} (HsP) & \smamath{fpr} \\
        \midrule
        DT & 1.0000 & 1.0000 & 0.0016 & 1.0000 & 0.0000 & 0.0003 \\
        RF & 0.9984 & 0.0000 & 0.0008 & 1.0000 & 0.0000 & 0.0000 \\
        HGB & 0.9984 & 0.0000 & 0.0016 & 1.0000 & 0.0000 & 0.0003 \\
        SVM & 0.9828 & 0.0000 & 0.0408 & 0.9946 & 0.0000 & 0.0188 \\
        DNN & 0.9953 & 0.0000 & 0.0071 & 0.9995 & 0.0000 & 0.0038 \\
        \bottomrule
        \end{tabular}
    }
\end{table}

\clearpage

\section{Supplementary Material}
\noindent
The specific commands used for the experiments described in Appendix~\ref{sapp:attacks} are provided in Listing~\ref{lst:hsp}.

\begin{lstlisting}[
frame=single,
breaklines=true, 
basicstyle=\ttfamily\footnotesize,
caption=Commands used for our additional HsP-based attacks.,
label={lst:hsp},
basicstyle=\tiny,
belowskip=-5mm,
float=t
]
## wfuzz (fuzzying)
wfuzz -t 4 -s 1 -w wordlist/general/common.txt http://$IP/FUZZ
wfuzz -t 4 -X HEAD -L -w wordlist/general/common.txt http://$IP/FUZZ
wfuzz -t 4 -s 1 -w wordlist/general/common.txt http://$IP/FUZZ
wfuzz -t 4 -w wordlist/general/common.txt http://$IP/FUZZ
wfuzz -t 1 -w wordlist/general/common.txt http://$IP/FUZZ


## slowloris (ddos)
python3 ./slowloris.py $IP (default)
python3 ./slowloris.py $IP -s 150 -ua (randomize user-agents)
python3 ./slowloris.py $IP -s 150 --sleeptime 30
python3 ./slowloris.py $IP -s 50

## nmap (portscan)
nmap $IP -n -Pn --top-ports 100 -sA -T4 --max-retries 1
nmap $IP -n -p 79,80 -Pn -sT -T3
nmap $IP -n -Pn --top-ports 20 -sS -sV --version-intensity 7 -T3 --max-retries 1
nmap $IP -n -Pn --top-ports 100 -sT -T4 --max-retries 1
nmap $IP -n -Pn --top-ports 100 -sS -T3 --reason --max-retries 1
nmap $IP -n -Pn --top-ports 20 -sU -sV --version-light -T3 --max-retries 1
nmap $IP -n -p 79,80 -Pn -sS -sV --version-light



\end{lstlisting}

\end{document}